\documentclass[12pt]{article}
\usepackage{amsmath, amssymb}
\usepackage{geometry}
\usepackage{graphicx}
\usepackage{hyperref}
\usepackage{natbib}
\usepackage{setspace}
\usepackage{booktabs}
\usepackage{longtable}
\begin{document}

\title{Extending Prais–Winsten Regression to Panel Data with Higher-Order Autoregressive Errors: A Simulation Study}
\author{\small Ariel Linden, DrPH\vspace{-4pt}\\
\small University of California, San Francisco\vspace{-4pt}\\
\small Department\ of Medicine, Division of Clinical Informatics \& Digital Transformation (DoC-IT)\vspace{-4pt}\\
\small San Francisco, CA, USA}
\date{}
\maketitle

\begin{abstract}
\noindent
Panel and cross-sectional time-series data frequently exhibit
higher-order autoregressive serial dependence, but existing panel
estimators combining generalized least squares (GLS) with
panel-corrected standard errors (PCSEs) are limited to first-order
autoregressive errors. To address this limitation, we extend the Prais–Winsten AR(k) generalized least squares (GLS) transformation to panel data within the Beck–Katz panel-corrected standard error (PCSE) framework. This extension is implemented in the community-contributed Stata package \texttt{xtpraisk}. As the panel extension of Prais--Winsten, \texttt{xtpraisk} is the
natural comparator to \texttt{xtscc}---the panel extension of
Newey--West---representing parametric and nonparametric
approaches to macro panel serial dependence. We conduct a Monte Carlo simulation to validate the statistical properties of \texttt{xtpraisk} and compare its finite-sample performance with the Driscoll–Kraay estimator implemented in \texttt{xtscc}. The simulation spans
autoregressive orders 1 through 3, three autocorrelation scenarios
(mild positive, oscillatory, and high persistent), three panel sizes,
six series lengths, and five effect sizes, with 2,000 replications per
condition and six performance measures. \texttt{xtpraisk} consistently achieved higher power than \texttt{xtscc} across all conditions, reflecting genuine GLS efficiency rather than anticonservative inference---confirmed by standard error ratios near 1.0 throughout. \texttt{xtscc} exhibited systematic
standard error underestimation and inflated Type~I error at short series
lengths, with both deficiencies worsening with AR order; these are
finite-sample phenomena that resolve as the number of time periods
increases. Both methods were essentially unbiased. Misspecification of
the AR order did not degrade \texttt{xtpraisk}'s inferential performance,
and cross-panel correlation and panel size had negligible effects on the relative performance of either estimator. Across the conditions examined, \texttt{xtpraisk} consistently provided better-calibrated inference and higher statistical power than \texttt{xtscc}. These findings suggest that \texttt{xtpraisk} may be advantageous when valid inference and statistical efficiency are both priorities, particularly at short series lengths, higher AR orders, and under persistent autocorrelation.

\end{abstract}


\section{Introduction}

Panel and cross-sectional time-series (CSTS) data---repeated
observations on fixed units such as countries, states, regions,
hospitals, or industries---are widely used across the social,
health, and policy sciences \citep{hsiao2003, baltagi2021}. This data
structure arises in a broad range of empirical settings, including
comparative political economy \citep{krieger2022}, social policy
research \citep{jacquesetal2024}, and public health
\citep{beyliketal2022}. CSTS data also underpin several
quasi-experimental evaluation designs. One prominent example is the
multiple-group (controlled) interrupted time-series design, in which
treated and comparison units are observed repeatedly before and after a
policy intervention \citep{linden2015}. The availability of repeated
pre-intervention observations also helps distinguish intervention-related
changes from regression to the mean \citep{linden2007rtm}. These applications often fall
within the panel structure emphasized by \citet{beckkatz1995}: a
relatively small to moderate number of units (approximately 10--100)
observed over comparatively long time horizons, often comprising 20--50
or more repeated observations.

The temporal and cross-sectional structure of these data presents
several challenges for statistical inference that go beyond those
encountered in single time-series analysis. Within each unit, errors
are often serially correlated over time. Across units, errors may be
heteroskedastic---varying in variance from one unit to another---and
contemporaneously correlated, meaning that units subject to common
shocks or shared environments may exhibit correlated errors at the same
point in time. Single time-series estimators such as Prais--Winsten
\citep{prais1954} and Newey--West \citep{neweywest1987} are designed to
address within-series serial correlation, but they do not account for
the between-unit heteroskedasticity and contemporaneous correlation
that characterize panel data. Ignoring any of these features can
produce incorrect standard errors and unreliable hypothesis tests
\citep{beckkatz1995, driscollkraay1998}.

Two widely used approaches for addressing all three complications
simultaneously are the estimators proposed by \citet{beckkatz1995} and
\citet{driscollkraay1998}. \citet{beckkatz1995} demonstrated that
feasible generalized least squares (FGLS) combined with
panel-corrected standard errors (PCSEs) provides reliable inference
across a broad range of panel structures. In this framework, a
Prais--Winsten generalized least squares (GLS) transformation
\citep{prais1954} removes within-unit serial correlation, while the
PCSE sandwich estimator accounts for remaining heteroskedasticity and
contemporaneous correlation across units. This approach is implemented
in Stata through the official \texttt{xtpcse} command
\citep{statacorp2023}. A key limitation of the Beck--Katz framework,
however, is its assumption that errors follow a first-order
autoregressive (AR(1)) process. Higher-order autoregressive dependence
is common in empirical panel data but cannot be accommodated directly
within the standard implementation.

\citet{driscollkraay1998} proposed an alternative that avoids
specifying a particular autoregressive structure. Their estimator
extends the Newey--West \citep{neweywest1987} heteroskedasticity- and
autocorrelation-consistent (HAC) covariance matrix estimator from the
single time-series setting to panel data by applying the nonparametric
correction to cross-sectional moment averages rather than individual
unit series. The resulting standard errors are robust to within-unit
serial correlation, cross-unit heteroskedasticity, and contemporaneous
correlation of arbitrary form. Because no assumptions are required
regarding the order of the underlying autoregressive process, the
estimator has become a popular alternative for panel time-series
analysis. In Stata, this approach is implemented through the
community-contributed \texttt{xtscc} command \citep{hoechle2007}.

Despite the widespread use of these two estimators, an important gap
remains. Both Prais--Winsten and Newey--West have been extended from
the single time-series setting to panel data---the former through the
Beck--Katz framework and the latter through Driscoll--Kraay---but the
two extensions are not symmetric in their treatment of serial
dependence. The Driscoll--Kraay estimator accommodates autoregressive
errors of arbitrary order nonparametrically, whereas the Beck--Katz
framework is restricted to AR(1). Consequently, no existing panel
estimator accommodates AR($k > 1$) errors within the GLS-plus-PCSE
framework. \citet{vougas2021} extended the Prais--Winsten algorithm to
AR($k$) processes of arbitrary order for single time series, and
\citet{linden2026a, linden2026b} subsequently implemented and evaluated
this extension in the multiple-group interrupted time-series setting.
The community-contributed Stata package \texttt{xtpraisk}
\citep{linden2026c} implements this extension for panel data, filling a
gap in the available panel time-series estimators. It combines the
exact Prais--Winsten GLS transformation for AR($k$) errors with
panel-specific Yule--Walker estimation of the autoregressive parameter
vector and the Beck--Katz PCSE sandwich estimator. When $k = 1$,
\texttt{xtpraisk} reproduces the Beck--Katz estimator---and therefore
\texttt{xtpcse}---exactly. When $k > 1$, it extends the framework to
accommodate higher-order serial dependence while preserving the
familiar PCSE approach to inference. When serial dependence extends
beyond a single lag, researchers must therefore choose between a
correctly specified nonparametric estimator and a misspecified
parametric one; \texttt{xtpraisk} is designed to eliminate this
trade-off.

The introduction of \texttt{xtpraisk} naturally invites comparison
with \texttt{xtscc}. Both estimators are designed for the same class
of panel data and address the same inferential challenges, but they
rely on fundamentally different strategies. \texttt{xtpraisk}
explicitly models the AR($k$) error process, removes serial dependence
through a GLS transformation, and then applies PCSEs to the
transformed data. In contrast, \texttt{xtscc} makes no parametric
assumptions about the autocorrelation structure, instead correcting for
serial dependence nonparametrically through weighted lagged
cross-sectional moment conditions. When the AR($k$) specification is
correct, the GLS approach should yield more efficient coefficient
estimates and better-calibrated standard errors than a nonparametric
correction. Whether this theoretical advantage translates into superior
finite-sample performance across the range of autoregressive structures
and panel dimensions encountered in applied research is the central
empirical question motivating this study.

The comparison is also substantively important because it contrasts two alternative approaches to handling serial dependence that occupy the same methodological niche. Whereas \texttt{xtpraisk} extends the Prais–Winsten generalized least squares framework to higher-order autoregressive processes in panel data, \texttt{xtscc} extends the Newey–West heteroskedasticity- and autocorrelation-consistent framework to the same setting. The comparison therefore isolates the consequences of adopting a parametric versus nonparametric strategy for serial dependence, rather than differences arising from estimators developed for distinct methodological objectives. The comparison therefore isolates the consequences of adopting a parametric versus nonparametric strategy for serial dependence, rather than differences arising from estimators developed for distinct methodological objectives. Alternative panel estimators often target different inferential settings or panel asymptotics and were therefore beyond the scope of the present investigation.

This comparison also extends our previous simulation work.
\citet{linden2026a} documented a power--inference trade-off between
analogous parametric and nonparametric approaches in the AR(1)
multiple-group interrupted time-series setting, while \citet{linden2026b}
showed that this trade-off becomes more pronounced under AR(2) and
AR(3) error processes. Whether the same pattern holds in panel
data---where PCSEs replace unit-specific standard errors and
cross-sectional dependence introduces an additional layer of
complexity---remains unknown.

To address this question, we conduct a Monte Carlo simulation study
comparing \texttt{xtpraisk} and \texttt{xtscc} across a fully crossed
design spanning autoregressive orders 1 through 3, three
autocorrelation scenarios for each order (mild positive, oscillatory,
and highly persistent), three panel sizes ($N \in \{10, 15, 20\}$),
six series lengths ($T \in \{10, 20, 30, 50, 75, 100\}$), and five
effect sizes, including a null condition. Following \citet{burton2006},
we evaluate six performance metrics: statistical power, Type I error
rate, 95\% confidence interval coverage, percentage bias, root mean
squared error (RMSE), and the standard error (SE) ratio. Additional
misspecification and sensitivity analyses complement the primary
results.

The remainder of the paper is organized as follows. Section~2 describes
the methods. Section~3 presents the simulation results. Section~4 provides an illustrative numerical example based on a prediabetes disease management study to demonstrate the practical consequences of estimator choice.
Section~5 discusses the findings, and Section~6 concludes.


\section{Methods}

\subsection{The generalized panel data model}

The standard panel data regression model takes the form
\begin{equation}
  y_{it} = \mathbf{x}_{it}'\boldsymbol{\beta} + \varepsilon_{it},
  \quad i = 1, \ldots, N, \quad t = 1, \ldots, T,
  \label{eq:panelmodel}
\end{equation}
where $y_{it}$ is the outcome for unit $i$ at time $t$;
$\mathbf{x}_{it}$ is a $q \times 1$ vector of regressors, including a
constant; $\boldsymbol{\beta}$ is the corresponding vector of
coefficients; and $\varepsilon_{it}$ is the error term. The errors are
assumed to follow an autoregressive process of order $k$, AR($k$):
\begin{equation}
  \varepsilon_{it} = \rho_1 \varepsilon_{i,t-1} + \rho_2 \varepsilon_{i,t-2}
    + \cdots + \rho_k \varepsilon_{i,t-k} + u_{it},
  \label{eq:ARk}
\end{equation}
where $u_{it} \sim \mathrm{i.i.d.}(0, \sigma^2_u)$ and
$\rho_j$ is the partial autocorrelation coefficient at lag $j$, for
$j = 1, 2, \ldots, k$ \citep{hamilton1994, wooldridge2020}.
When $k = 1$, equation~(\ref{eq:ARk}) reduces to the familiar AR(1)
process $\varepsilon_{it} = \rho\,\varepsilon_{i,t-1} + u_{it}$.

In addition to within-unit serial correlation, panel data errors may
exhibit two further complications. Errors may be \emph{panel
heteroskedastic}, meaning that $\mathrm{Var}(\varepsilon_{it}) =
\sigma^2_i$ differs across units. They may also be
\emph{contemporaneously correlated}, meaning that
$\mathrm{Cov}(\varepsilon_{it}, \varepsilon_{jt}) = \sigma_{ij} \neq 0$
for $i \neq j$ at the same time point $t$. Following \citet{beckkatz1995},
we denote the $N \times N$ matrix of contemporaneous covariances as
$\boldsymbol{\Sigma}$, with typical element $\sigma_{ij}$. The full
$NT \times NT$ error covariance matrix is
$\boldsymbol{\Omega} = \boldsymbol{\Sigma} \otimes \mathbf{I}_T$
after the within-unit autocorrelation has been removed by the
Prais--Winsten transformation (see Section~\ref{sec:xtpraisk}).

\subsection{Estimation approaches}

\subsubsection{The Driscoll--Kraay estimator}
\label{sec:dkest}

\citet{driscollkraay1998} showed that the Newey--West
\citep{neweywest1987} heteroskedasticity- and auto\-correlation-consistent
(HAC) covariance matrix estimator, when applied to the cross-sectional
averages of the moment conditions rather than to individual unit series,
yields standard errors that are robust to within-unit serial correlation,
cross-unit heteroskedasticity, and contemporaneous correlation of
arbitrary form. The key insight is that averaging the moment conditions
over units reduces the dimensionality of the covariance problem from
$NR \times NR$ to $R \times R$, where $R$ is the number of orthogonality
conditions, so that the estimator remains feasible regardless of the
size of $N$ \citep{driscollkraay1998}.

For the linear panel model in equation~(\ref{eq:panelmodel}), the
ordinary least squares (OLS) estimator of $\boldsymbol{\beta}$ is:
\begin{equation}
  \hat{\boldsymbol{\beta}} = \left(\mathbf{X}'\mathbf{X}\right)^{-1}
    \mathbf{X}'\mathbf{y},
  \label{eq:OLS}
\end{equation}
where $\mathbf{X}$ and $\mathbf{y}$ stack all $NT$ observations. The
Driscoll--Kraay covariance matrix estimator is:
\begin{equation}
  \widehat{\mathrm{Var}}_{\mathrm{DK}}\!\left(\hat{\boldsymbol{\beta}}\right)
    = \left(\mathbf{X}'\mathbf{X}\right)^{-1}
      \hat{S}_T
      \left(\mathbf{X}'\mathbf{X}\right)^{-1},
  \label{eq:DKvar}
\end{equation}
where $\hat{S}_T$ is a nonparametric HAC estimator of the long-run
covariance matrix of the cross-sectional averages of the score,
$\bar{h}_t = N^{-1}\sum_{i=1}^{N} \mathbf{x}_{it}\hat{\varepsilon}_{it}$,
with $\hat{\varepsilon}_{it}$ the OLS residuals. Specifically,
\begin{equation}
  \hat{S}_T = \hat{\Gamma}(0) +
    \sum_{j=1}^{m} w_j \left[\hat{\Gamma}(j) + \hat{\Gamma}(j)'\right],
  \label{eq:DKS}
\end{equation}
where $\hat{\Gamma}(j) = T^{-1}\sum_{t=j+1}^{T}
\bar{h}_t \bar{h}_{t-j}'$ is the sample autocovariance of the
cross-sectional score averages at lag $j$, and $w_j = 1 -
j/(m+1)$ are the Bartlett (triangular) kernel weights
\citep{neweywest1987}. The bandwidth $m$ is selected using the
data-driven rule of \citet{neweywest1994}:
\begin{equation}
  m = \left\lfloor 4\left(\frac{T}{100}\right)^{2/9} \right\rfloor.
  \label{eq:bandwidth}
\end{equation}
This approach is implemented in Stata through the
community-contributed \texttt{xtscc} command \citep{hoechle2007}.

\subsubsection{The xtpraisk estimator}
\label{sec:xtpraisk}

\texttt{xtpraisk} is a feasible generalized least squares (FGLS)
estimator for panel data with AR($k$) errors. It extends the
Prais--Winsten \citep{prais1954} GLS transformation to AR($k$)
processes following \citet{vougas2021}, combines it with the
panel-by-panel Yule--Walker estimation approach of
\citet{parkmitchell1980}, and applies the PCSE sandwich of
\citet{beckkatz1995} to the transformed data.

\paragraph{Statistical model.}
Let $k$ denote the AR lag order and $\boldsymbol{\rho} =
(\rho_1, \ldots, \rho_k)'$ the vector of AR parameters. The model
consists of equations~(\ref{eq:panelmodel}) and~(\ref{eq:ARk}).
The regression coefficients $\boldsymbol{\beta}$ and the AR parameter
vector $\boldsymbol{\rho}$ are estimated jointly by the iterative GLS
algorithm described below.

\paragraph{Stationarity.}
The stationarity condition for AR($k$) requires that all eigenvalues
of the $k \times k$ companion matrix $\mathbf{C}$ have modulus strictly
less than 1 \citep{hamilton1994}, where
\begin{equation}
  \mathbf{C} =
  \begin{pmatrix}
    \rho_1 & \rho_2 & \cdots & \rho_{k-1} & \rho_k \\
    1      & 0      & \cdots & 0          & 0      \\
    0      & 1      & \cdots & 0          & 0      \\
    \vdots &        & \ddots & \vdots     & \vdots \\
    0      & 0      & \cdots & 1          & 0
  \end{pmatrix}\!.
  \label{eq:companion}
\end{equation}
For AR(1) this reduces to $|\rho| < 1$; for AR(2) to the triangle
conditions $\rho_1 + \rho_2 < 1$, $\rho_2 - \rho_1 < 1$, and
$|\rho_2| < 1$ \citep{hamilton1994}.

\paragraph{Yule--Walker estimation.}
Starting from OLS residuals $\hat{u}_{it} = y_{it} -
\mathbf{x}_{it}'\hat{\boldsymbol{\beta}}_{\mathrm{OLS}}$, the AR
parameters are estimated via the pooled Yule--Walker normal equations.
Cross-product matrices are accumulated within each panel and summed
across panels, yielding a single pooled estimate of
$\boldsymbol{\rho}$ \citep{parkmitchell1980}. For $k = 1$ this
gives the standard moment estimator
\begin{equation}
  \hat{\rho} = \frac{\sum_{i=1}^{N}\sum_{t=2}^{T}
    \hat{u}_{it}\hat{u}_{i,t-1}}
    {\sum_{i=1}^{N}\sum_{t=2}^{T} \hat{u}_{i,t-1}^2}.
  \label{eq:YW1}
\end{equation}
For $k = 2$ the $2 \times 2$ system
\begin{equation}
  \begin{pmatrix}
    \sum \hat{u}_{i,t-1}^2 & \sum \hat{u}_{i,t-1}\hat{u}_{i,t-2} \\
    \sum \hat{u}_{i,t-1}\hat{u}_{i,t-2} & \sum \hat{u}_{i,t-2}^2
  \end{pmatrix}
  \begin{pmatrix} \rho_1 \\ \rho_2 \end{pmatrix}
  =
  \begin{pmatrix}
    \sum \hat{u}_{it}\hat{u}_{i,t-1} \\
    \sum \hat{u}_{it}\hat{u}_{i,t-2}
  \end{pmatrix}
  \label{eq:YW2}
\end{equation}
is solved via LU decomposition \citep{golubvanloan1996}. For AR($k >
2$) the general $k \times k$ system $\mathbf{A}\boldsymbol{\rho} =
\mathbf{b}$ is formed analogously and solved in the same way.

\paragraph{Prais--Winsten GLS transformation.}
Given $\hat{\boldsymbol{\rho}}$, the exact Prais--Winsten GLS
transformation is applied independently to each panel. For
observations $t = k+1, \ldots, T$ within panel $i$, the AR($k$)
filter yields the quasi-differenced quantities
\begin{equation}
  \tilde{y}_{it} = y_{it} - \sum_{j=1}^{k} \hat{\rho}_j\, y_{i,t-j},
  \quad
  \tilde{\mathbf{x}}_{it} = \mathbf{x}_{it} -
    \sum_{j=1}^{k} \hat{\rho}_j\, \mathbf{x}_{i,t-j}.
  \label{eq:filter}
\end{equation}
The first $k$ observations of each panel are transformed using the
upper-left $k \times k$ block of the Cholesky factor of
$\mathbf{V}_k^{-1}$, where $\mathbf{V}_k$ is the $k \times k$
autocovariance matrix of the AR($k$) process. This matrix is computed
analytically using the closed-form expression of
\citet{galbraith1974}. For $k = 1$ this reduces to the familiar
Prais--Winsten weight $\sqrt{1 - \rho^2}$ applied to the first
observation \citep{prais1954}. For $k \geq 2$ the procedure applies
to a $k \times k$ system, with no special-case treatment required
\citep{vougas2021}. The transformation is restarted at the beginning
of every panel, and cross-panel autocorrelation is assumed to be zero.

\paragraph{GLS estimation and PCSE sandwich.}
Let $\tilde{\mathbf{y}}_i$ and $\tilde{\mathbf{X}}_i$ denote the
Prais--Winsten transformed outcome and regressor matrix for panel $i$.
Stacking across panels gives $\tilde{\mathbf{y}}$ and
$\tilde{\mathbf{X}}$. The GLS coefficient estimator is:
\begin{equation}
  \hat{\boldsymbol{\beta}} =
    \left(\tilde{\mathbf{X}}'\tilde{\mathbf{X}}\right)^{-1}
    \tilde{\mathbf{X}}'\tilde{\mathbf{y}}.
  \label{eq:GLS}
\end{equation}
Following \citet{beckkatz1995}, the PCSE sandwich estimator of the
covariance matrix of $\hat{\boldsymbol{\beta}}$ is:
\begin{equation}
  \widehat{\mathrm{Var}}_{\mathrm{PCSE}}\!\left(\hat{\boldsymbol{\beta}}\right)
    = \left(\tilde{\mathbf{X}}'\tilde{\mathbf{X}}\right)^{-1}
      \left(\tilde{\mathbf{X}}'\hat{\boldsymbol{\Omega}}
      \tilde{\mathbf{X}}\right)
      \left(\tilde{\mathbf{X}}'\tilde{\mathbf{X}}\right)^{-1},
  \label{eq:PCSE}
\end{equation}
where $\hat{\boldsymbol{\Omega}} = \hat{\boldsymbol{\Sigma}} \otimes
\mathbf{I}_T$ and $\hat{\boldsymbol{\Sigma}}$ is the $N \times N$
matrix of estimated contemporaneous cross-panel covariances, with
typical element $\hat{\sigma}_{ij} = T^{-1}\sum_{t=1}^{T}
\tilde{e}_{it}\tilde{e}_{jt}$, where $\tilde{e}_{it} = \tilde{y}_{it}
- \tilde{\mathbf{x}}_{it}'\hat{\boldsymbol{\beta}}$ are the
transformed residuals.

\paragraph{Iterative algorithm.}
The algorithm alternates between Yule--Walker estimation of
$\boldsymbol{\rho}$ (Step 1) and GLS estimation of
$\boldsymbol{\beta}$ (Step 2) until convergence, declared when
$\max_j |\hat{\rho}_j^{(\mathrm{new})} - \hat{\rho}_j^{(\mathrm{old})}|
< \tau$, where $\tau = 10^{-6}$ by default \citep{judge1985,
vougas2021}. The PCSE sandwich in equation~(\ref{eq:PCSE}) is computed
once at convergence. When $k = 1$, \texttt{xtpraisk} replicates the
\citet{beckkatz1995} estimator---and therefore \texttt{xtpcse}---exactly.

\subsection{Simulation strategy}
\label{sec:simstrat}

Table~1 presents the simulation inputs. The primary
objective was to evaluate whether, and to what extent, the AR order
and autocorrelation structure differentially influence the
finite-sample performance of \texttt{xtpraisk} and \texttt{xtscc}
across a range of panel dimensions, series lengths, and effect sizes.

The data-generating process (DGP) followed model~(\ref{eq:panelmodel})
with $\boldsymbol{\beta} = (\beta_0, \beta_1)'$, where $\beta_0 = 10$
is the intercept and $\beta_1$ is the slope coefficient of primary
interest. The single continuous regressor $x_{it}$ was drawn
independently from a standard normal distribution. The random error
$u_{it}$ in equation~(\ref{eq:ARk}) was drawn from $\mathrm{N}(0, 1)$.

For each AR order ($k = 1, 2, 3$), three autocorrelation scenarios
were specified to represent qualitatively distinct patterns of serial
dependence plausible in applied panel data:
\begin{enumerate}
  \item \emph{Mild positive autocorrelation}: AR(1) $\rho = 0.4$;
    AR(2) $\boldsymbol{\rho} = (0.4, 0.2)$; AR(3)
    $\boldsymbol{\rho} = (0.4, 0.2, 0.1)$. The autocovariance
    function decays monotonically, producing the kind of smooth
    positive serial dependence commonly encountered in routine panel
    data. The maximum companion matrix eigenvalue is approximately
    0.40, 0.69, and 0.72 for AR(1), AR(2), and AR(3), respectively.
  \item \emph{Oscillatory autocorrelation}: AR(1) $\rho = -0.4$;
    AR(2) $\boldsymbol{\rho} = (0.5, -0.4)$; AR(3)
    $\boldsymbol{\rho} = (0.7, -0.3, 0.15)$. The negative
    coefficient produces an autocovariance function that alternates
    in sign, creating a complex spectral structure. The maximum
    companion matrix eigenvalue is approximately 0.40, 0.63, and 0.84
    for AR(1), AR(2), and AR(3), respectively.
  \item \emph{High persistent positive autocorrelation}: AR(1)
    $\rho = 0.7$; AR(2) $\boldsymbol{\rho} = (0.7, 0.2)$; AR(3)
    $\boldsymbol{\rho} = (0.6, 0.25, 0.1)$. The autocovariance
    function decays very slowly, approaching but remaining within
    the unit circle. The maximum companion matrix eigenvalue is
    approximately 0.70, 0.90, and 0.93 for AR(1), AR(2), and AR(3),
    respectively.
\end{enumerate}

The initial AR errors were drawn from their stationary distributions.
For AR(1), the stationary variance is $\sigma^2_u / (1 - \rho^2)$.
For AR(2), the $2 \times 2$ stationary covariance matrix
$\mathbf{V}_2$ was computed using the closed-form expression of
\citet{galbraith1974}. For AR(3), a burn-in period of 200 periods
was used to initialize the process. Panel sizes of $N \in \{10, 15,
20\}$ and series lengths of $T \in \{10, 20, 30, 50, 75, 100\}$ were
examined. Effect sizes for $\beta_1$ were set to $\{0, 0.25, 0.50,
0.75, 1.00\}$, where $\beta_1 = 0$ provides the null condition for
Type I error evaluation. The treatment was set to a single binary
predictor, with the proportion of treated observations fixed at 0.5.

Primary results are presented for $N = 10$ and all three AR orders.
Each figure displays results for a single performance measure, with
\texttt{xtpraisk} in the left column and \texttt{xtscc} in the right
column; rows represent effect sizes and lines distinguish autocorrelation
scenarios. AR(1) and AR(3) figures are presented in the Supplement;
AR(2) figures appear in the main text as the central higher-order case.
Supplementary analyses for misspecification and sensitivity follow the
same layout. To isolate the role of panel size, a final set of analyses
examined Type~I error and SE ratio under AR(2) errors across
$N \in \{10, 15, 20\}$, holding effect size fixed at $\Delta = 0.20$
and allowing autocorrelation scenario and series length to vary.

\subsection{Misspecification analysis}
\label{sec:misspec}

A supplementary simulation examined the consequences of AR order
underspecification---arguably the most common form of model
misspecification in practice \citep{linden2026b}. Data were generated
under an AR(2) process (using the same three autocorrelation scenarios
as the primary analysis) but the model was estimated assuming AR(1).
This scenario reflects the realistic situation in which a researcher
defaults to a single-lag correction without conducting formal
autocorrelation diagnostics, or in which software defaults (such as
\texttt{xtpcse}) impose an AR(1) structure. Panel sizes ($N \in \{10,
15, 20\}$), series lengths ($T \in \{10, 20, 30, 50, 75, 100\}$), and
effect sizes ($\beta_1 \in \{0, 0.20\}$) followed the same ranges as
the primary simulations. Both \texttt{xtpraisk} and \texttt{xtscc}
were evaluated under the misspecified AR(1) model, and performance was
assessed using the same six metrics as in the primary analysis.

\subsection{Sensitivity analysis}
\label{sec:sensitivity}

The sensitivity analysis examined whether the primary findings are
robust to a higher degree of contemporaneous cross-panel correlation.
In the primary simulations, the regressor $x_{it}$ is drawn
independently across units, producing only modest cross-panel
dependence in the errors. In the sensitivity analysis, a common factor
structure was introduced: $x_{it} = \lambda f_t + v_{it}$, where
$f_t \sim \mathrm{N}(0,1)$ is a common factor, $v_{it} \sim
\mathrm{N}(0,1)$ is a unit-specific idiosyncratic component, and
$\lambda \in \{0.5, 1.0\}$ controls the strength of cross-panel
dependence. This factor structure is consistent with the Monte Carlo
designs used by \citet{driscollkraay1998} and \citet{hoechle2007}
to evaluate the finite-sample performance of the Driscoll--Kraay
estimator. All other inputs followed the primary simulation design.

\subsection{Performance measures}
\label{sec:perf}

The performance of \texttt{xtpraisk} and \texttt{xtscc} was evaluated
using six metrics following \citet{burton2006}:
\begin{enumerate}
  \item \emph{Statistical power} ($1 - \beta$): the proportion of
    replications in which the null hypothesis $H_0\colon \beta_1 = 0$
    is correctly rejected at $\alpha = 0.05$.
  \item \emph{Type I error rate}: the proportion of replications in
    which $H_0\colon \beta_1 = 0$ is incorrectly rejected when
    $\beta_1 = 0$ (null condition), evaluated at $\alpha = 0.05$.
  \item \emph{95\% confidence interval (CI) coverage}: the proportion
    of replications in which the nominal 95\% CI contains the true
    value of $\beta_1$.
  \item \emph{Percentage bias}: $100 \times
    (\bar{\hat{\beta}}_1 - \beta_1) / \beta_1$, where
    $\bar{\hat{\beta}}_1$ is the mean of the estimated coefficient
    across replications.
  \item \emph{Root mean squared error (RMSE)}:
    $\sqrt{R^{-1}\sum_{r=1}^{R}(\hat{\beta}_{1r} - \beta_1)^2}$,
    where $R = 2{,}000$ is the number of replications.
  \item \emph{Standard error (SE) ratio}: the mean model-based
    standard error of $\hat{\beta}_1$ across replications divided by
    the empirical standard deviation of $\hat{\beta}_1$ across
    replications. A ratio of 1.0 indicates perfect SE calibration;
    values below 1.0 indicate underestimation of variability
    (anticonservative inference); values above 1.0 indicate
    overestimation (conservative inference). This measure directly
    explains the mechanistic link between SE calibration and observed
    differences in Type I error and coverage.
\end{enumerate}
In total, the primary simulation encompassed 810 unique design conditions
(3 AR orders $\times$ 3 autocorrelation scenarios $\times$ 3 panel sizes
$\times$ 6 series lengths $\times$ 5 effect sizes). Combined with 108
misspecification conditions (3 autocorrelation scenarios $\times$ 3 panel
sizes $\times$ 6 series lengths $\times$ 2 effect sizes) and 120 sensitivity
conditions, 1,038 unique design conditions were evaluated in total, each
replicated 2,000 times, yielding 2,076,000 simulated datasets. All hypothesis tests used two-sided Wald tests at $\alpha = 0.05$, and all analyses were conducted using Stata version 19.0 \citep{statacorp2023}.


\section{Results}

\subsection{Power}

Figure~1 presents statistical power ($1 - \beta$) for \texttt{xtpraisk}
and \texttt{xtscc} under AR(2) error structures across three effect
sizes ($\Delta \in \{0.10, 0.20, 0.30\}$) and three autocorrelation
scenarios. Results for AR(1) and AR(3) are presented in
Appendix~Figures~A1 and A2, respectively.

Under AR(2) errors, \texttt{xtpraisk} accumulated power at a consistent
rate across all three autocorrelation scenarios, with the curves tracking
closely together throughout. At $\Delta = 0.10$, near-maximum power was
achieved by approximately $T = 75$--100; at $\Delta = 0.20$ and $\Delta
= 0.30$, convergence occurred by $T \approx 30$--40 and $T \approx 20$,
respectively. \texttt{xtscc} showed substantially lower power across all
conditions, with performance diverging markedly by autocorrelation
scenario. Under mild positive and oscillatory autocorrelation,
\texttt{xtscc} accumulated power at a moderate rate, approaching but not
reaching maximum power by $T = 100$ at $\Delta = 0.10$. Under high
persistent autocorrelation, power was severely suppressed: at $\Delta =
0.10$ it barely exceeded 0.35 at $T = 100$, and even at $\Delta = 0.20$
it reached only approximately 0.85 by the end of the series length range.

The pattern under AR(1) (Appendix Figure~A1) was broadly similar, though
the gap between estimators was smaller. \texttt{xtscc} continued to
accumulate power across the full range of $T$ under all three scenarios,
reaching approximately 0.60 at $T = 100$ for the high persistent scenario
at $\Delta = 0.10$. Under AR(3) (Appendix Figure~A2), the divergence
between estimators was most extreme. \texttt{xtpraisk} maintained its
consistent power accumulation across scenarios, though the high persistent
scenario was noticeably slower at $\Delta = 0.10$, reaching approximately
0.75 at $T = 100$. For \texttt{xtscc}, the high persistent scenario
exhibited virtually no power accumulation across the full $T$ range at
$\Delta = 0.10$, remaining near 0.20--0.25 throughout, and the
oscillatory scenario displayed a non-monotone dip near $T = 20$ before
recovering. Only under mild positive autocorrelation did \texttt{xtscc}
accumulate power at a meaningful rate under AR(3).


\subsection{95\% Confidence Interval Coverage}

Figure~2 presents 95\% confidence interval coverage for \texttt{xtpraisk}
and \texttt{xtscc} under AR(2) error structures. Results for AR(1) and
AR(3) are presented in Appendix~Figures~A3 and A4, respectively.

Under AR(2) errors, \texttt{xtpraisk} maintained near-nominal coverage
throughout, with all three autocorrelation scenarios clustering tightly
between approximately 93\% and 96\% across the full range of series
lengths. No systematic trend was evident with increasing $T$, and
coverage was well-calibrated from the shortest series lengths examined.
\texttt{xtscc} showed a markedly different pattern: coverage began
substantially below nominal at short series lengths, reaching
approximately 87\%--92\% at $T = 10$, with the high persistent scenario
starting lowest at around 87\%. Coverage improved steadily with
increasing $T$, approaching but not fully reaching nominal levels by
$T = 100$, where all three scenarios settled in the range of 93\%--95\%.
The pattern was consistent across effect sizes.

The same contrast was observed under AR(1) (Appendix Figure~A3),
though with a smaller initial deficit for \texttt{xtscc}: coverage
began near 90\%--91\% at $T = 10$ and rose to approximately 93\%--95\%
by $T = 100$. \texttt{xtpraisk} again maintained near-nominal coverage
throughout. Under AR(3) (Appendix Figure~A4), the initial deficit for
\texttt{xtscc} was largest, with coverage starting near 84\%--86\% at
$T = 10$ before climbing to approximately 93\%--94\% by $T = 100$.
\texttt{xtpraisk} remained well-calibrated across all AR(3) conditions,
though with slightly more fluctuation at short series lengths, including
a small dip in the oscillatory scenario around $T = 20$--30.


\subsection{Type I Error}

Figure~3 presents Type I error rates for \texttt{xtpraisk} and
\texttt{xtscc} across all three AR orders. Unlike the power and coverage
figures, all AR orders are presented in a single combined figure, with
rows representing AR order and columns representing estimator.

\texttt{xtpraisk} maintained well-controlled Type I error rates
throughout, hovering between approximately 4\% and 7\% across all AR
orders, autocorrelation scenarios, and series lengths. Minor fluctuation
around the nominal 5\% level was present at short series lengths but
diminished with increasing $T$. No systematic inflation was observed
under any condition.

\texttt{xtscc} showed elevated Type I error at short series lengths
under all AR orders, with the degree of inflation increasing with AR
order. Under AR(1), rates began at approximately 8\%--11\% at $T = 10$
and declined to near-nominal levels by $T \approx 30$--40, where they
remained for the rest of the series length range. Under AR(2), the
initial elevation was somewhat higher (approximately 11\%--12\% at $T =
10$) but declined steeply, reaching near-nominal levels by $T \approx
40$--50. Under AR(3), inflation was most pronounced, with rates reaching
13\%--14\% at $T = 10$ before declining. Convergence to near-nominal
levels was slower than under AR(1) or AR(2), with some residual
elevation persisting through $T \approx 50$--75 before settling near 5\%
by $T = 100$.


\subsection{Percentage Bias}

Figure~4 presents percentage bias for \texttt{xtpraisk} and
\texttt{xtscc} under AR(2) error structures. Results for AR(1)
and AR(3) are presented in Appendix~Figures~A5 and A6, respectively.

Both estimators were essentially unbiased across all AR orders,
autocorrelation scenarios, and series lengths. At larger effect sizes
($\Delta = 0.20$ and $\Delta = 0.30$), bias remained negligible for
both estimators throughout the full range of $T$, with all scenarios
clustering tightly around zero. At the smallest effect size ($\Delta =
0.10$), modest instability appeared at very short series lengths,
particularly for \texttt{xtscc}, which showed larger transient
deviations at $T = 10$ that quickly resolved by $T = 20$. The magnitude
of this instability increased with AR order: under AR(1) and AR(2) the
deviations at $T = 10$ were generally within $\pm$5--8\%, whereas under
AR(3) \texttt{xtscc} exhibited a spike of approximately 25\% in one
scenario at $T = 10$ before collapsing to near zero by the next series
length. \texttt{xtpraisk} showed smaller and more stable fluctuations
at short series lengths across all AR orders. Beyond $T = 20$, both
estimators were indistinguishable from zero under all conditions.


\subsection{Root Mean Squared Error}

Figure~5 presents root mean squared error (RMSE) for \texttt{xtpraisk}
and \texttt{xtscc} under AR(2) error structures. Results for AR(1) and
AR(3) are presented in Appendix~Figures~A7 and A8, respectively.

\texttt{xtpraisk} produced closely overlapping RMSE profiles across
all three autocorrelation scenarios under AR(2), declining monotonically
from approximately 0.10 at $T = 10$ to approximately 0.03 at $T = 100$.
This pattern was essentially identical across all AR orders and effect
sizes, indicating that \texttt{xtpraisk}'s estimation precision is
largely insensitive to the autocorrelation structure once the AR order
is correctly specified. \texttt{xtscc} showed higher RMSE at short
series lengths, with the degree of elevation depending strongly on the
autocorrelation scenario. Under mild positive and oscillatory
autocorrelation, \texttt{xtscc} started at approximately 0.12--0.15 at
$T = 10$ and declined to values comparable to \texttt{xtpraisk} by
$T = 100$. Under high persistent autocorrelation, the initial elevation
was considerably larger, reaching approximately 0.21 at $T = 10$, and
the gap relative to \texttt{xtpraisk} persisted throughout, with
\texttt{xtscc} remaining at approximately 0.07 at $T = 100$ compared
to 0.03 for \texttt{xtpraisk}.

The same pattern held under AR(1) (Appendix Figure~A7), though with
smaller initial differences between estimators: \texttt{xtscc} began
at approximately 0.11--0.14 at $T = 10$ and the three scenarios tracked
closely together throughout. Under AR(3) (Appendix Figure~A8), the
divergence between estimators was most extreme. \texttt{xtpraisk}
remained stable near 0.10 at $T = 10$ declining to 0.03 at $T = 100$,
while \texttt{xtscc} under high persistent autocorrelation started near
0.28 at $T = 10$ and, despite a steep decline, remained near 0.08--0.09
at $T = 100$. The mild positive and oscillatory scenarios for
\texttt{xtscc} under AR(3) converged to values near \texttt{xtpraisk}
by $T = 100$.


\subsection{Standard Error Ratio}

Figure~6 presents the standard error (SE) ratio for \texttt{xtpraisk}
and \texttt{xtscc} under AR(2) error structures. Results for AR(1) and
AR(3) are presented in Appendix~Figures~A9 and A10, respectively.

The SE ratio results provide the mechanistic explanation for the
coverage and Type~I error patterns reported above. \texttt{xtpraisk}
maintained ratios close to 1.0 throughout, with all three
autocorrelation scenarios fluctuating between approximately 0.94 and
1.05 across the full range of series lengths under AR(2). This
near-perfect calibration was consistent across all AR orders and
effect sizes, confirming that \texttt{xtpraisk}'s standard errors
accurately reflect the true sampling variability of the coefficient
estimates under all conditions examined.

\texttt{xtscc} exhibited systematic SE underestimation at short series
lengths under all AR orders, with the degree of underestimation
increasing with AR order. Under AR(2), ratios began at approximately
0.75--0.78 at $T = 10$, rising steeply to approximately 0.94--1.00 by
$T = 75$--100. Under AR(1) (Appendix Figure~A9), the initial deficit
was smaller, with ratios starting near 0.85 at $T = 10$ and recovering
to approximately 0.93--0.95 by $T = 100$. Under AR(3) (Appendix
Figure~A10), the underestimation was most severe, with ratios beginning
near 0.70--0.75 at $T = 10$, rising to approximately 0.94--0.98 by
$T = 100$. Across all AR orders, the three autocorrelation scenarios
tracked closely together for \texttt{xtscc}, and the pattern was
essentially identical across effect sizes, indicating that the SE
miscalibration is a structural property of the HAC estimator in short
panels rather than a sample-size or effect-size artifact.


\subsection{Effect of Panel Size}
\label{sec:Neffect}

Figures~9 and~10 present Type~I error and SE ratio results under AR(2)
errors for $N \in \{10, 15, 20\}$, holding the effect size fixed at
$\Delta = 0.20$. Rows correspond to autocorrelation scenarios; curves
distinguish panel sizes.

For \texttt{xtpraisk}, Type~I error remained near-nominal and SE ratios
remained near 1.0 across all three values of $N$, all autocorrelation
scenarios, and the full range of series lengths, with the three curves
essentially indistinguishable throughout. For \texttt{xtscc}, the
small-$T$ inflation documented in the primary analyses was present at
all three panel sizes, and the rate of convergence toward nominal as $T$
increased was similarly unaffected by $N$. The three curves for
\texttt{xtscc} overlapped closely in both figures across all
autocorrelation scenarios.


\subsection{Misspecification Analysis}

Figure~7 presents Type~I error and power for \texttt{xtpraisk} and
\texttt{xtscc} when data are generated under an AR(2) process but both
estimators are fitted with a lag(1) model.

Type~I error remained well-controlled for \texttt{xtpraisk} under
misspecification, with all three autocorrelation scenarios fluctuating
between approximately 4\% and 8\% across the full range of series
lengths---comparable to the correctly specified results reported in
Section~3.3. \texttt{xtscc} showed modest initial elevation of
approximately 9\%--11\% at $T = 10$, declining quickly to near-nominal
levels by $T \approx 20$--30 and remaining there thereafter. Both
estimators were therefore relatively robust to AR order
underspecification with respect to Type~I error control.

Power under misspecification showed a more pronounced contrast.
\texttt{xtpraisk} retained high power despite the underspecified lag
order, with all three scenarios converging to near-maximum power by
$T \approx 30$--40, beginning from approximately 0.50--0.61 at $T =
10$. \texttt{xtscc} showed substantially lower power throughout, with
the mild positive and oscillatory scenarios approaching maximum power
by $T \approx 75$--100 but the high persistent scenario reaching only
approximately 0.85 at $T = 100$.


\subsection{Sensitivity Analysis}

Figure~8 presents Type~I error and 95\% confidence interval coverage
for \texttt{xtpraisk} and \texttt{xtscc} under AR(2) errors with a
common factor inducing cross-panel correlation, for factor loadings
$\lambda = 0.5$ and $\lambda = 1.0$ under the mild positive and high
persistent autocorrelation scenarios.

Cross-panel correlation at either level of $\lambda$ did not
meaningfully alter the inferential behavior of either estimator.
\texttt{xtpraisk} maintained near-nominal Type~I error and coverage
across all conditions, with no degradation relative to the primary
results. \texttt{xtscc} exhibited the same small-$T$ inflation observed
in the primary analyses, converging toward nominal as $T$ increased.
Increasing $\lambda$ from 0.5 to 1.0 had negligible additional effect
on either method, and the pattern was consistent across both
autocorrelation scenarios.


\section{Illustrative Numerical Example}

\subsection{Background and Study Design}

The simulation study establishes the finite-sample operating characteristics of \texttt{xtpraisk} and \texttt{xtscc} under a wide range of panel dimensions, autoregressive structures, and effect sizes. To complement those results, we present a numerical illustration demonstrating the practical consequences of estimator choice in a realistic panel-data setting.

Because publicly available panel datasets with independently verifiable higher-order autoregressive error structures are uncommon, and because the true error-generating process is necessarily unknown in empirical applications, the objective of this example is not methodological validation. Rather, it is to illustrate how different approaches to handling serial dependence can lead to different substantive conclusions when higher-order autoregressive processes are present.

Disease management programs target individuals with chronic conditions or elevated risk through structured behavioral and clinical interventions \citep{lindenadler2008, kullgren2018}. Prediabetes is a common target given the effectiveness of lifestyle interventions in preventing progression to type 2 diabetes \citep{biuso2007}. To provide a realistic empirical setting, the illustration is based on a hypothetical disease management study in which patients with prediabetes are followed longitudinally across multiple health systems.

The artificial study involves ten US regional health systems whose
patients with prediabetes were tracked monthly for 30 months. The first
15 months established a baseline average fasting blood glucose level,
after which all ten health systems simultaneously implemented a
coordinated lifestyle intervention program at month~16. All systems
were monitored for a further 15 months following the intervention. The
outcome is the monthly average fasting blood glucose level (mg/dL)
aggregated across patients within each health system, yielding a
balanced panel of $N = 10$ units and $T = 30$ time periods.

Data were generated in Stata using a standard panel regression
data-generating process: $y_{it} = \beta_0 + \beta_1 x_{it} + u_{it}$,
where $x_{it}$ is a post-intervention indicator (0 pre, 1 post) common
to all units, $\beta_0 = 108$~mg/dL (within the clinical prediabetes
range of 100--125~mg/dL), and $\beta_1 = -2.5$~mg/dL, reflecting a
modest but clinically meaningful reduction in average blood glucose
following the intervention. A standard deviation of 3~mg/dL was applied
to reflect realistic month-to-month variability within health systems.
High persistent positive autocorrelation was specified to reflect
plausible serial dependence in longitudinal aggregate glucose
measurements. Two datasets were generated using identical parameters,
differing only in AR order: AR(1) ($\rho = 0.7$) and AR(2)
($\boldsymbol{\rho} = (0.7, 0.2)$). Each dataset was analyzed using
both \texttt{xtpraisk} and \texttt{xtscc}, with the lag order matched
to the AR order of the generating process. Because the data-generating
process is identical across estimators, the fitted values do not differ
between methods; they diverge only in their standard errors and
inferential conclusions.

\subsection{Results}

Table~2 presents the intervention effect estimates, standard errors,
95\% confidence intervals, and $p$-values for each AR order and
estimation method. These results represent a single realization;
coefficient estimates will vary across replications, though the
simulation study confirms both methods are approximately unbiased on
average.

Under AR(1), both methods agreed: the intervention was associated with
a substantial and statistically significant reduction in average blood
glucose of approximately 4.1~mg/dL, with similar standard errors and
the same clinical conclusion. Under AR(2), the methods reached opposite
conclusions. \texttt{xtpraisk} returned a coefficient of
$-1.36$~mg/dL with a standard error of 0.83, yielding $p = 0.103$ and
a confidence interval that included zero---correctly signalling
insufficient evidence given the persistent higher-order autocorrelation
structure. \texttt{xtscc}, by contrast, returned the same coefficient
but with a standard error of only 0.44, yielding $p = 0.026$ and a
confidence interval entirely below zero. The SE underestimation by
\texttt{xtscc}---a ratio of 0.44 to 0.83, less than half---directly
reflects the finite-sample HAC miscalibration documented in the
simulations at this series length and AR order. A disease management
program could be recommended for broad implementation based on
\texttt{xtscc}'s false confidence, while \texttt{xtpraisk} correctly
signals that the evidence is insufficient. It should be noted that
with a longer study window, both methods would likely produce
comparably robust results, as the simulation findings show
\texttt{xtscc}'s inferential performance converges toward
\texttt{xtpraisk}'s as $T$ increases.

\section{Discussion}

The fundamental finding of this study is that the trade-off between
power and inferential validity documented for analogous time-series
estimators under AR(1) \citep{linden2026a} and higher-order
\citep{linden2026b} autoregressive errors not only persists in the panel
setting but is resolved in a qualitatively different way. In the
multiple-group interrupted time-series context, OLS with Newey--West
standard errors achieved higher power at the cost of inflated Type~I
error and poor coverage, with deficiencies that worsened as AR order and
series length increased \citep{linden2026b}. In the panel context
examined here, \texttt{xtpraisk} achieves higher power than
\texttt{xtscc} while simultaneously maintaining well-calibrated
inference. The two estimators are not simply better or worse versions of
each other; they differ in both their treatment of serial dependence and
in how the panel dimension modifies that treatment. Table~3
summarizes performance across AR orders.

\subsection{Statistical Power}

The power advantage of \texttt{xtpraisk} over \texttt{xtscc} is a
genuine efficiency gain rather than a consequence of anticonservative
inference. This distinguishes the present findings from those reported
for OLS-NW versus Prais--Winsten in the time-series setting
\citep{linden2026a, linden2026b}, where OLS-NW's apparent power
advantage was attributable to inflated false positive rates. The power
advantage is therefore consistent with the efficiency gain from correctly specified GLS estimation, consistent with asymptotic theory
\citep{judge1985, wooldridge2020}.

The magnitude of the advantage varied predictably with AR order and autocorrelation scenario. It was largest under high persistent autocorrelation and at higher AR orders, conditions under which GLS-based approaches would be expected to yield the greatest variance reduction relative to OLS-based alternatives. Under oscillatory autocorrelation, the gap narrowed, suggesting that these error structures may be less favorable to the GLS transformation than highly persistent positive processes.

\subsection{Coverage and Type~I Error}

The coverage and Type~I error results reveal a pattern that differs fundamentally from the time-series setting. In prior work, the inferential deficiencies of OLS-Newey–West increased rather than diminished with series length under persistent autocorrelation \citep{linden2026b}. No analogous pattern emerged in the present panel simulations. Instead, the Type~I error inflation and coverage deficits observed for \texttt{xtscc} were concentrated at short series lengths and steadily diminished as $T$ increased.

This pattern is consistent with the asymptotic properties of the Driscoll–Kraay estimator \citep{driscollkraay1998}. Although \texttt{xtscc} exhibited inflated Type~I error rates and below-nominal coverage when series were short, both measures converged toward their nominal levels as the number of time periods increased. The practical implication is that the inferential deficiencies of \texttt{xtscc} are primarily a finite-sample concern, whereas \texttt{xtpraisk} maintained near-nominal coverage and Type~I error control throughout the full range of conditions examined.

\subsection{Bias, RMSE, and SE Ratio}

Both methods were essentially unbiased across all conditions, confirming that the strict exogeneity of the simulated covariate was preserved under all data-generating processes. Differences in performance therefore arise from variance estimation and inferential calibration rather than systematic error in point estimation. This finding parallels those reported for the time-series setting \citep{linden2026a,linden2026b} and suggests that the panel extension does not introduce additional bias.

\texttt{xtpraisk} produced substantially lower RMSE than \texttt{xtscc} at short series lengths, with the gap largest under high persistent autocorrelation and higher AR orders. This pattern is consistent with the efficiency gains expected from GLS estimation. By removing serial dependence prior to estimation, \texttt{xtpraisk} reduces the effective variance of the coefficient estimator, whereas \texttt{xtscc} retains the OLS coefficient estimates and adjusts only the estimated covariance matrix. As $T$ increased, RMSE converged across methods, consistent with the asymptotic equivalence of consistent estimators under standard regularity conditions.

The SE ratio results provide a useful explanation for the coverage and Type~I error patterns reported above. \texttt{xtpraisk} maintained ratios near 1.0 throughout, indicating that its model-based standard errors closely tracked the empirical sampling variability of the coefficient estimates. This finding is consistent with the Beck–Katz framework, in which the PCSE sandwich is applied after the GLS transformation has removed within-unit serial dependence, leaving the between-unit covariance structure to be estimated \citep{beckkatz1995}.

In contrast, \texttt{xtscc} exhibited systematic SE underestimation at short series lengths, with the degree of underestimation increasing as AR order increased. A plausible explanation is that the nonparametric HAC correction becomes increasingly difficult to estimate accurately when the autocorrelation structure is more complex and the number of time periods is limited. The fact that the underestimation diminished as $T$ increased, but worsened with AR order, suggests that the finite-sample performance of \texttt{xtscc} depends strongly on the amount of information available to estimate the long-run covariance structure. These findings are consistent with the observed patterns in coverage, Type~I error, and power.

\subsection{Misspecification, Sensitivity, and Panel Size}

The misspecification results provide practical reassurance for
\texttt{xtpraisk} users. Fitting an AR(1) model when the true process
is AR(2) did not meaningfully degrade \texttt{xtpraisk}'s inferential
performance, with Type~I error remaining near-nominal throughout. This
robustness is consistent with the GLS transformation's known resilience
to moderate lag order underspecification \citep{judge1985}, and mirrors
the analogous finding from the time-series setting
\citep{linden2026b}. For \texttt{xtscc}, misspecification produced
results indistinguishable from the correctly specified case, as expected
given that its nonparametric correction does not depend on lag order
specification.

The sensitivity analysis confirmed that cross-panel correlation induced
by a common factor does not alter the relative performance of the two
estimators. \texttt{xtpraisk}'s PCSE sandwich is designed to capture
contemporaneous correlation across units \citep{beckkatz1995}, and the
simulation results confirm that it does so effectively across both
levels of factor loading examined here. \texttt{xtscc} also proved
robust to cross-panel correlation at both $\lambda$ values, consistent
with the Driscoll--Kraay estimator's theoretical accommodation of
cross-sectional dependence \citep{driscollkraay1998}.

The panel size results reinforce the same conclusion. Performance
differences between the two estimators are driven by $T$ rather than
$N$: neither estimator's inferential behavior was meaningfully affected
by increasing the number of panels from 10 to 20, under any
autocorrelation scenario or series length examined.

\subsection{Practical Recommendations}

The results support a clear practical recommendation. \texttt{xtpraisk}
is the preferred estimator across the full range of conditions examined:
it provides better-calibrated inference than \texttt{xtscc} at short to
moderate series lengths, higher power at all series lengths, and lower
RMSE. The advantage is most pronounced when series are short, AR order
is high, and autocorrelation is persistent---precisely the conditions
most commonly encountered in applied panel time-series research
\citep{beckkatz1995, hsiao2003}. Researchers working with $T < 30$
should be particularly cautious with \texttt{xtscc}, as Type~I error
inflation and SE underestimation are most severe in this range.

When $k = 1$, \texttt{xtpraisk} reproduces \texttt{xtpcse} exactly,
so existing workflows built around \texttt{xtpcse} require no
adjustment. When higher-order dependence is plausible---as is often the
case with monthly, weekly, or daily panel data---\texttt{xtpraisk} with
$k > 1$ offers a straightforward extension that preserves the familiar
PCSE framework while accommodating the richer autocorrelation structure.
The misspecification results further suggest that researchers need not
identify the exact AR order: fitting a slightly underspecified model
incurs negligible inferential cost for \texttt{xtpraisk}. Regardless of estimator, researchers should assess the autocorrelation
structure of their data, report the lag order used, and consider
sensitivity analyses under alternative specifications \citep{lindenroberts2005}.

\subsection{Limitations}

Several limitations of the present study should be noted. The
simulations were restricted to balanced panels, a single covariate with
strict exogeneity, normally distributed innovations, and a common AR
parameter across units. These restrictions may not reflect the full
range of applied panel data structures. The factor model used in the
sensitivity analysis represents one approach to cross-panel dependence;
other forms of spatial or network dependence were not examined. The
maximum series length was $T = 100$, and some conditions under high
persistent autocorrelation had not fully converged at this horizon. The
maximum AR order examined was 3; whether the PCSE sandwich remains
well-calibrated and \texttt{xtscc}'s small-$T$ inflation persists at
AR(4) and beyond is unknown. Finally, only two estimators were evaluated. This choice was intentional, as the objective of the study was to compare parametric and nonparametric approaches to handling serial dependence within the same macro-panel inferential framework. Other approaches to panel dependence may address different methodological objectives and were therefore beyond the scope of the present investigation. Future research could extend these comparisons to additional classes of panel estimators that accommodate higher-order autoregressive dynamics.

\section{Conclusion}

The trade-off between power and inferential validity documented for
analogous time-series estimators \citep{linden2026a, linden2026b}
persists in the macro panel setting but takes a qualitatively different
form. \texttt{xtpraisk} achieves higher power than \texttt{xtscc}
across all AR orders and autocorrelation scenarios examined, and this
advantage reflects genuine GLS efficiency rather than anticonservative
inference---a distinction confirmed by SE ratios near 1.0 throughout.
\texttt{xtscc}'s inferential deficiencies are a finite-sample
phenomenon, resolving as $T$ increases, rather than the structural
deterioration observed for OLS-NW under persistent autocorrelation in
prior work \citep{linden2026b}. Both methods were essentially unbiased,
confirming that performance differences are about variance estimation,
not point accuracy. Across the conditions examined, \texttt{xtpraisk} consistently provided better-calibrated inference, higher statistical power, and lower RMSE than \texttt{xtscc}. These findings suggest that \texttt{xtpraisk} may be advantageous when valid inference and statistical efficiency are both priorities, particularly at short series lengths, higher AR orders, and under persistent autocorrelation. Future research should examine estimator performance under non-Gaussian outcomes, unbalanced panels, heterogeneous AR parameters across units, and higher-order serial dependence structures beyond AR(3).

\clearpage
\bibliographystyle{plainnat}
\bibliography{xtpraisk_refs}

\clearpage
\section*{Abbreviations}

\noindent AR: Autoregressive; FGLS: Feasible generalized least squares;
GLS: Generalized least squares; HAC: Heteroskedasticity- and
autocorrelation-consistent standard errors; OLS: Ordinary least squares;
PCSE: Panel-corrected standard errors; RMSE: Root mean squared error;
SE: Standard error; TSCS: Time-series cross-sectional data.

\section*{Supplementary Information}

\noindent The Supplement contains figures for AR(1) and AR(3) power,
coverage, bias, RMSE, and SE ratio (Appendix Figures A1--A10). Stata
code used in this paper is found at:
\url{https://github.com/ariellinden/xtpraisk}

\section*{Authors' Contributions}

\noindent AL conceived the study and its design, conducted all analyses,
wrote the manuscript, and takes public responsibility for its content.

\section*{Funding}

\noindent There was no funding associated with this work.

\section*{Ethics Approval and Consent to Participate}

\noindent Not applicable.

\section*{Consent for Publication}

\noindent Not applicable.

\section*{Competing Interests}

\noindent The author declares no competing interests.

\section*{Acknowledgements}

\noindent I am grateful to Dimitrios V. Vougas for graciously providing
the MATLAB code that served as the basis for implementation of the
\texttt{xtpraisk} package.

\clearpage

\parbox{14cm}{\raggedright\textbf{Table 1.} Simulation design and inputs.}\vspace{2pt}
\begin{longtable}{p{6.8cm} p{7.2cm}}
\label{tab:inputs} \\
\toprule
\textbf{Parameter} & \textbf{Value / Description} \\
\midrule
\endfirsthead
\multicolumn{2}{l}{\footnotesize\textit{Table~1 continued}} \\
\toprule
\textbf{Parameter} & \textbf{Value / Description} \\
\midrule
\endhead
\midrule
\multicolumn{2}{r}{\footnotesize\textit{Continued on next page}} \\
\endfoot
\bottomrule
\multicolumn{2}{p{14cm}}{%
\vspace{4pt}
\footnotesize
\textit{Note}: AR = autoregressive; DGP = data-generating process; GLS = generalized
least squares; PCSE = panel-corrected standard errors; RMSE = root mean squared error;
SE = standard error. $\rho$ structures for AR(1) follow \citet{linden2026a}; AR(2) and
AR(3) follow \citet{linden2026b}. Primary simulation results are shown for $N = 10$;
supplementary figures show $N = 15$ and $N = 20$. The effect size $\beta_1 = 0.05$
was simulated but is omitted from all figures due to uniformly low power across all
conditions and AR orders.} \\
\endlastfoot
\multicolumn{2}{l}{\textit{Regression model}} \\
Model
  & $y_{it} = \beta_0 + \beta_1 x_{it} + u_{it}$ \\
Error process
  & $u_{it} = \rho_1 u_{i,t-1} + \cdots + \rho_k u_{i,t-k} + \varepsilon_{it}$,
    \quad $\varepsilon_{it} \sim \mathrm{iid}\,\mathcal{N}(0,1)$ \\
Covariate
  & $x_{it} \sim \mathrm{iid}\,\mathcal{N}(0,1)$, redrawn every replication
    (strict exogeneity) \\
Intercept ($\beta_0$) & 10 \\
Innovation SD ($\sigma$) & 1 \\
\midrule
\multicolumn{2}{l}{\textit{Effect sizes ($\beta_1$)}} \\
Null (Type I error)  & 0 \\
Small                & 0.05 \\
Medium               & 0.10 \\
Large                & 0.20 \\
Very large           & 0.30 \\
\midrule
\multicolumn{2}{l}{\textit{Panel design}} \\
Number of panels ($N$)       & 10, 15, 20 \\
Time periods per panel ($T$) & 10, 20, 30, 50, 75, 100 \\
Panel balance                & Balanced ($N_i = T$ for all $i$) \\
AR parameter structure       & Common $\rho$ across panels (pooled) \\
\midrule
\multicolumn{2}{l}{\textit{Autoregressive order and scenarios}} \\
AR(1) --- Scenario 1 (mild positive)   & $\rho = 0.4$ \\
AR(1) --- Scenario 2 (\mbox{oscillatory})     & $\rho = -0.4$ \\
AR(1) --- Scenario 3 (high persistent) & $\rho = 0.7$ \\
AR(2) --- Scenario 1 (mild positive)   & $\boldsymbol{\rho} = (0.4,\; 0.2)$ \\
AR(2) --- Scenario 2 (\mbox{oscillatory})     & $\boldsymbol{\rho} = (0.5,\; -0.4)$ \\
AR(2) --- Scenario 3 (high persistent) & $\boldsymbol{\rho} = (0.7,\; 0.2)$ \\
AR(3) --- Scenario 1 (mild positive)   & $\boldsymbol{\rho} = (0.4,\; 0.2,\; 0.1)$ \\
AR(3) --- Scenario 2 (\mbox{oscillatory})     & $\boldsymbol{\rho} = (0.7,\; -0.3,\; 0.15)$ \\
AR(3) --- Scenario 3 (high persistent) & $\boldsymbol{\rho} = (0.6,\; 0.25,\; 0.1)$ \\
\midrule
\multicolumn{2}{l}{\textit{AR error initialization}} \\
AR(1)
  & $u_0 \sim \mathcal{N}(0,\;\sigma^2/(1-\rho^2))$ [stationary variance] \\
AR(2)
  & First 2 obs.\ from stationary distribution via \citet{galbraith1974}
    $\mathbf{V}_2$ matrix \\
AR(3)
  & 200-period burn-in; first 200 obs.\ discarded per panel \\
\midrule
\multicolumn{2}{l}{\textit{Estimators compared}} \\
\texttt{xtpraisk}
  & Prais--Winsten AR($k$) GLS with panel-corrected standard errors;
    lag specified to match true AR order \\
\texttt{xtscc}
  & Driscoll--Kraay HAC standard errors applied to OLS; lag specified
    to match \texttt{xtpraisk} at each condition \\
\midrule
\multicolumn{2}{l}{\textit{Misspecification analysis}} \\
Design    & DGP $=$ AR(2); both estimators fitted with lag(1) \\
$N$       & 10, 15, 20 \\
$T$       & 10, 20, 30, 50, 75, 100 \\
$\beta_1$ & 0 (Type I error), 0.20 (power) \\
\midrule
\multicolumn{2}{l}{\textit{Sensitivity analysis (cross-panel correlation)}} \\
DGP
  & AR(2) with common factor: $\varepsilon_{it} = \lambda f_t + \eta_{it}$,
    $f_t \sim \mathrm{iid}\,\mathcal{N}(0,1)$,
    $\eta_{it} \sim \mathrm{iid}\,\mathcal{N}(0,1)$ \\
Scenarios & Scenarios 1 (mild positive) and 3 (high persistent) only \\
$\lambda$ & 0.5 (moderate cross-panel correlation), 1.0 (strong) \\
$N$       & 10, 15, 20 \\
$T$       & 10, 20, 30, 50, 100 \\
$\beta_1$ & 0 (Type I error), 0.20 (power) \\
\midrule
\multicolumn{2}{l}{\textit{Performance measures \citep{burton2006}}} \\
Power
  & Proportion of replications rejecting $H_0\colon\beta_1=0$
    when $\beta_1 \neq 0$ \\
Type I error
  & Proportion of replications rejecting $H_0\colon\beta_1=0$
    when $\beta_1 = 0$ \\
95\% CI coverage
  & Proportion of replications in which the 95\% CI contains the
    true $\beta_1$ (evaluated when $\beta_1 \neq 0$) \\
Percentage bias  & $100 \times (\overline{\hat{\beta}}_1 - \beta_1)\,/\,\beta_1$ \\
RMSE             & $\sqrt{\mathrm{mean}(\hat{\beta}_1 - \beta_1)^2}$ \\
SE ratio
  & $\mathrm{mean(estimated\ SE)}\,/\,\mathrm{SD}(\hat{\beta}_1)$;
    values ${<}\,1$ indicate underestimation \\
\midrule
\multicolumn{2}{l}{\textit{Monte Carlo settings}} \\
Replications per condition    & 2,000 \\
Significance level ($\alpha$) & 0.05 (two-sided Wald test) \\
\end{longtable}


\clearpage

\begin{table}[h]
\begin{flushleft}
\parbox{12.5cm}{\raggedright\textbf{Table~2.} Illustrative example: intervention effect (mg/dL) by AR order and estimation method.}\\[4pt]
\label{tab:applied}
\begin{tabular}{llcccc}
\toprule
AR Order & Method & Coefficient & SE & $p$-value &
  \multicolumn{1}{c}{95\% CI} \\
\midrule
AR(1) & \texttt{xtpraisk} & $-4.07$ & $0.68$ & $<0.001$ & $(-5.40,\,-2.74)$ \\
      & \texttt{xtscc}    & $-4.26$ & $0.60$ & $<0.001$ & $(-5.62,\,-2.91)$ \\[4pt]
AR(2) & \texttt{xtpraisk} & $-1.36$ & $0.83$ & $0.103$  & $(-2.99,\;\;0.27)$ \\
      & \texttt{xtscc}    & $-1.17$ & $0.44$ & $0.026$  & $(-2.17,\,-0.18)$ \\
\bottomrule
\end{tabular}\\[4pt]
\parbox{12.5cm}{\footnotesize\textit{Note}: Both methods applied to the same dataset
at each AR order. SE\,=\,standard error; CI\,=\,confidence interval.
Under AR(1) both methods agree; under AR(2) \texttt{xtscc}
underestimates the standard error, producing a false positive while
\texttt{xtpraisk} correctly signals insufficient evidence.}
\end{flushleft}
\end{table}

\clearpage

{\setstretch{1.0}\small
\noindent\textbf{Table 3.} Summary comparison of simulation findings across AR(1), AR(2), and AR(3) error structures for \texttt{xtpraisk} and \texttt{xtscc} in panel time-series analysis.

\label{tab:summary}
\begin{longtable}{p{2.2cm} p{4.1cm} p{4.1cm} p{4.1cm}}
\toprule
\textbf{Performance Measure} & \textbf{AR(1)} & \textbf{AR(2)} & \textbf{AR(3)} \\
\midrule
\endfirsthead
\multicolumn{4}{l}{\footnotesize\textit{Table~3 continued}} \\
\toprule
\textbf{Performance Measure} & \textbf{AR(1)} & \textbf{AR(2)} & \textbf{AR(3)} \\
\midrule
\endhead
\midrule
\endfoot
\bottomrule
\multicolumn{4}{p{14.5cm}}{\footnotesize\textit{Note}: SE = standard error;
RMSE = root mean squared error; $T$ = number of time periods per panel.
All results shown for $N = 10$. \texttt{xtpraisk} uses Prais--Winsten AR($k$) GLS
with panel-corrected standard errors; \texttt{xtscc} uses Driscoll--Kraay HAC
standard errors applied to OLS\@. Lag order matched to the true AR order.} \\
\endlastfoot
Power
  & \texttt{xtpraisk} consistently higher than
    \texttt{xtscc}; advantage largest under high
    persistent autocorrelation. Reflects genuine GLS
    efficiency---SE ratios near 1.0 confirm
    well-calibrated inference throughout.
  & Pattern consistent with AR(1); power gap widens
    under high persistent autocorrelation.
    \texttt{xtpraisk} advantage confirmed as genuine
    efficiency gain across all scenarios.
  & Pattern consistent with AR(2); gap widens further.
    \texttt{xtpraisk} advantage most pronounced at
    short $T$ and remains substantial even at
    $T = 100$ under high persistent autocorrelation.
    \\
95\% Coverage
  & \texttt{xtpraisk} near-nominal throughout.
    \texttt{xtscc} below nominal at small $T$,
    recovering by $T \approx 50$.
  & \texttt{xtpraisk} near-nominal throughout.
    \texttt{xtscc} deficit more pronounced than AR(1);
    approaches nominal by $T \approx 75$.
  & \texttt{xtpraisk} near-nominal throughout.
    \texttt{xtscc} most severe deficit; approaches
    nominal only by $T \approx 75$--100 and remains
    below nominal under high persistent autocorrelation.
    \\
Type~I Error
  & \texttt{xtpraisk} near-nominal throughout.
    \texttt{xtscc} modestly inflated at small $T$;
    converges to nominal by $T \approx 30$--50.
  & \texttt{xtpraisk} near-nominal throughout.
    \texttt{xtscc} inflation more pronounced than
    AR(1); converges by $T \approx 50$--75.
  & \texttt{xtpraisk} near-nominal throughout.
    \texttt{xtscc} most pronounced inflation; under
    high persistent autocorrelation, approaches
    nominal only near $T = 100$.
    \\
Bias
  & Both methods essentially unbiased across all
    conditions and series lengths.
  & Both methods essentially unbiased; consistent
    with AR(1).
  & Both methods essentially unbiased; consistent
    with AR(1) and AR(2).
    \\
RMSE
  & \texttt{xtpraisk} substantially lower at small
    $T$; gap largest under high persistent
    autocorrelation. Both converge at large $T$.
  & \texttt{xtpraisk} advantage more pronounced than
    AR(1) at small $T$; convergence occurs at
    larger $T$.
  & \texttt{xtpraisk} advantage most pronounced of
    all AR orders; gap remains substantial under
    high persistent autocorrelation at moderate $T$.
    \\
SE Ratio
  & \texttt{xtpraisk} near 1.0 throughout.
    \texttt{xtscc} underestimates at small $T$
    (${\approx}0.85$ at $T = 10$); recovers to
    ${\approx}1.0$ by $T = 100$.
  & \texttt{xtpraisk} near 1.0 throughout.
    \texttt{xtscc} underestimation more severe than
    AR(1) (${\approx}0.75$--0.78 at $T = 10$);
    recovers by $T \approx 75$--100.
  & \texttt{xtpraisk} near 1.0 throughout.
    \texttt{xtscc} most severe underestimation
    (${\approx}0.70$--0.75 at $T = 10$); does not
    fully recover even at $T = 100$ under high
    persistent autocorrelation.
    \\
Misspecification
  & ---
  & Both methods robust to AR order
    underspecification. \texttt{xtpraisk} Type~I
    error near-nominal; \texttt{xtscc} shows same
    small-$T$ inflation as in correctly specified
    case. Power gap mirrors primary results.
  & Not examined.
    \\
\end{longtable}
}

\clearpage
\begin{figure}[p]
\centering
\includegraphics[width=\textwidth,height=0.85\textheight,keepaspectratio]{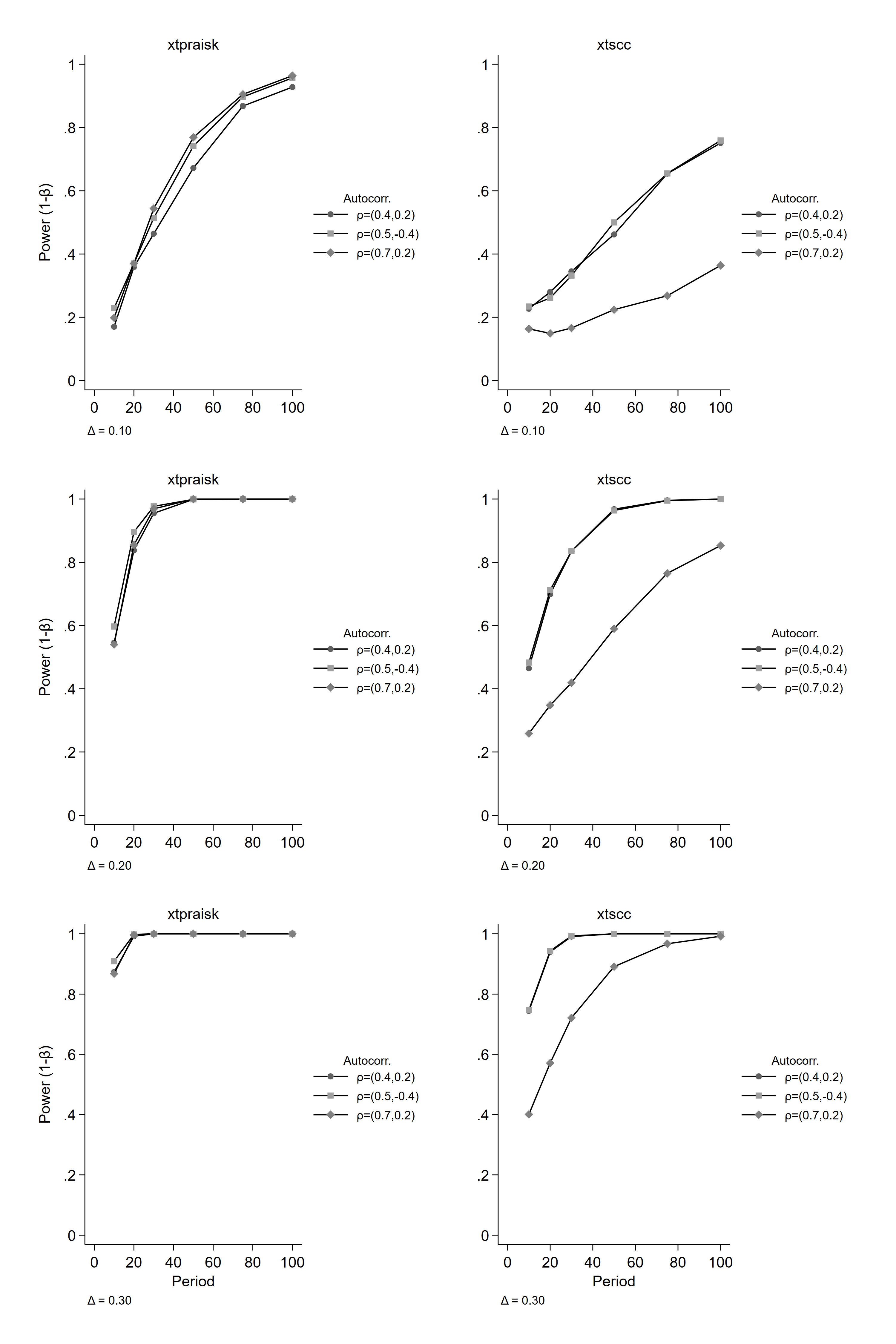}
\caption{Statistical power ($1-\beta$) for \texttt{xtpraisk} (left column) and \texttt{xtscc} (right column) under AR(2) error structures. Rows represent effect sizes ($\Delta = 0.10$, 0.20, 0.30). Lines distinguish autocorrelation scenarios: mild positive $\boldsymbol{\rho} = (0.4, 0.2)$ (circles); oscillatory $\boldsymbol{\rho} = (0.5, -0.4)$ (squares); high persistent $\boldsymbol{\rho} = (0.7, 0.2)$ (diamonds). $N = 10$.}
\label{fig:power_ar2}
\end{figure}

\clearpage
\begin{figure}[p]
\centering
\includegraphics[width=\textwidth,height=0.85\textheight,keepaspectratio]{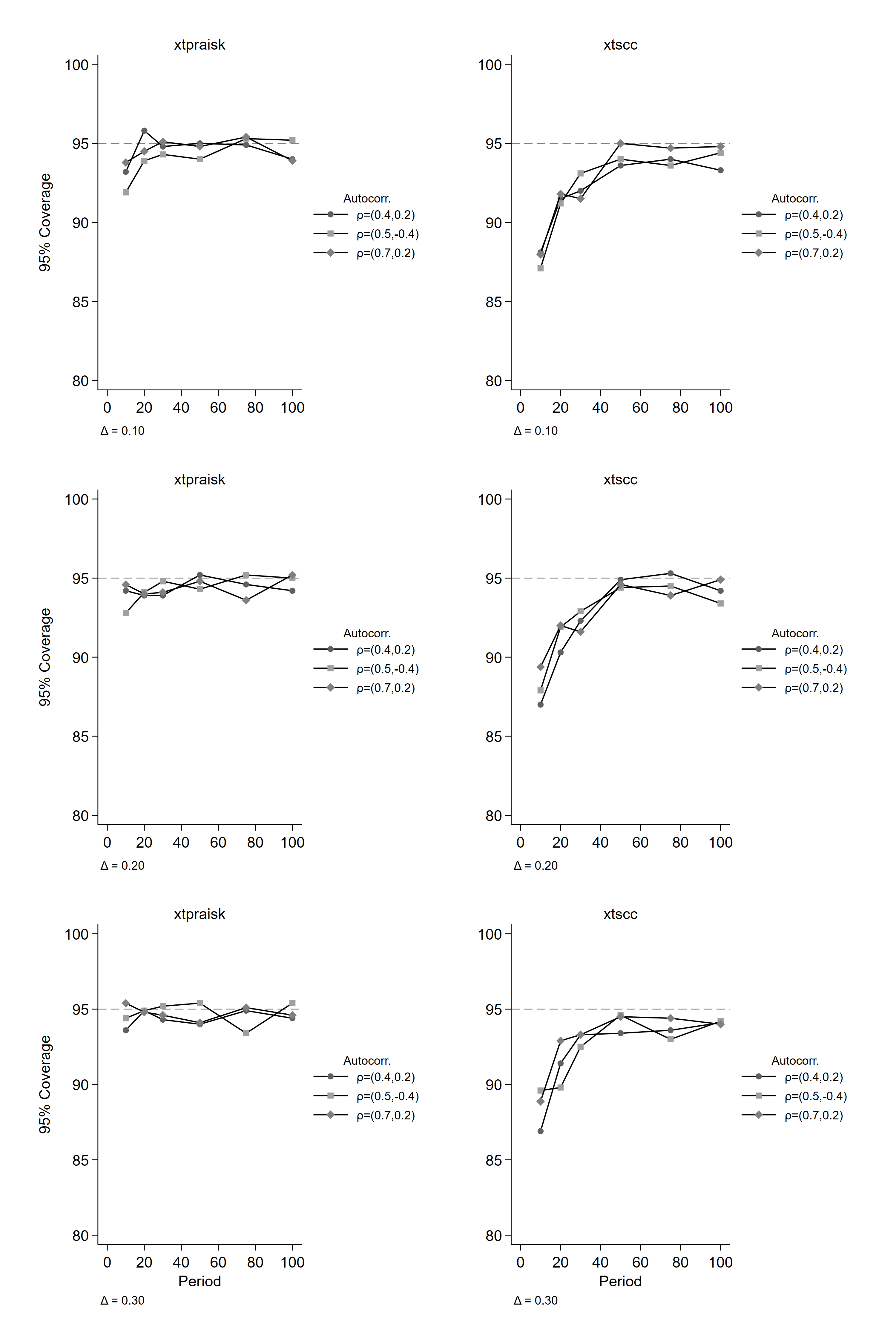}
\caption{95\% confidence interval coverage for \texttt{xtpraisk} (left column) and \texttt{xtscc} (right column) under AR(2) error structures. Rows represent effect sizes ($\Delta = 0.10$, 0.20, 0.30). Lines distinguish autocorrelation scenarios: mild positive $\boldsymbol{\rho} = (0.4, 0.2)$ (circles); oscillatory $\boldsymbol{\rho} = (0.5, -0.4)$ (squares); high persistent $\boldsymbol{\rho} = (0.7, 0.2)$ (diamonds). Dashed reference line at nominal 95\%. $N = 10$.}
\label{fig:coverage_ar2}
\end{figure}

\clearpage
\begin{figure}[p]
\centering
\includegraphics[width=\textwidth,height=0.85\textheight,keepaspectratio]{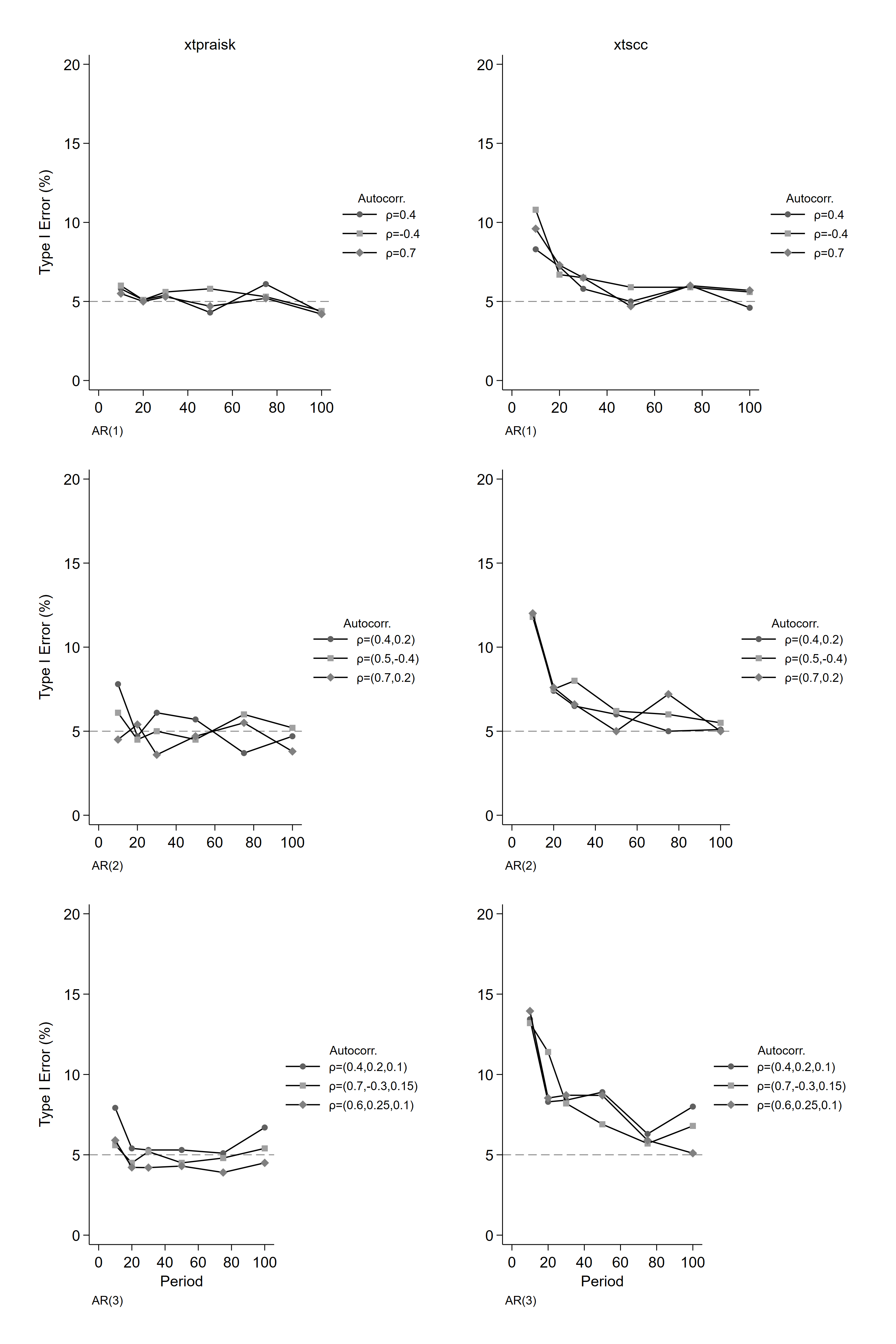}
\caption{Type~I error rates for \texttt{xtpraisk} (left column) and \texttt{xtscc} (right column) across AR orders 1 through 3 (rows). Lines distinguish autocorrelation scenarios: AR(1): mild positive $\rho = 0.4$ (circles), oscillatory $\rho = -0.4$ (squares), high persistent $\rho = 0.7$ (diamonds); AR(2): mild positive $\boldsymbol{\rho} = (0.4, 0.2)$ (circles), oscillatory $\boldsymbol{\rho} = (0.5, -0.4)$ (squares), high persistent $\boldsymbol{\rho} = (0.7, 0.2)$ (diamonds); AR(3): mild positive $\boldsymbol{\rho} = (0.4, 0.2, 0.1)$ (circles), oscillatory $\boldsymbol{\rho} = (0.7, -0.3, 0.15)$ (squares), high persistent $\boldsymbol{\rho} = (0.6, 0.25, 0.1)$ (diamonds). Dashed reference line at nominal 5\%. $N = 10$.}
\label{fig:type1}
\end{figure}

\clearpage
\begin{figure}[p]
\centering
\includegraphics[width=\textwidth,height=0.85\textheight,keepaspectratio]{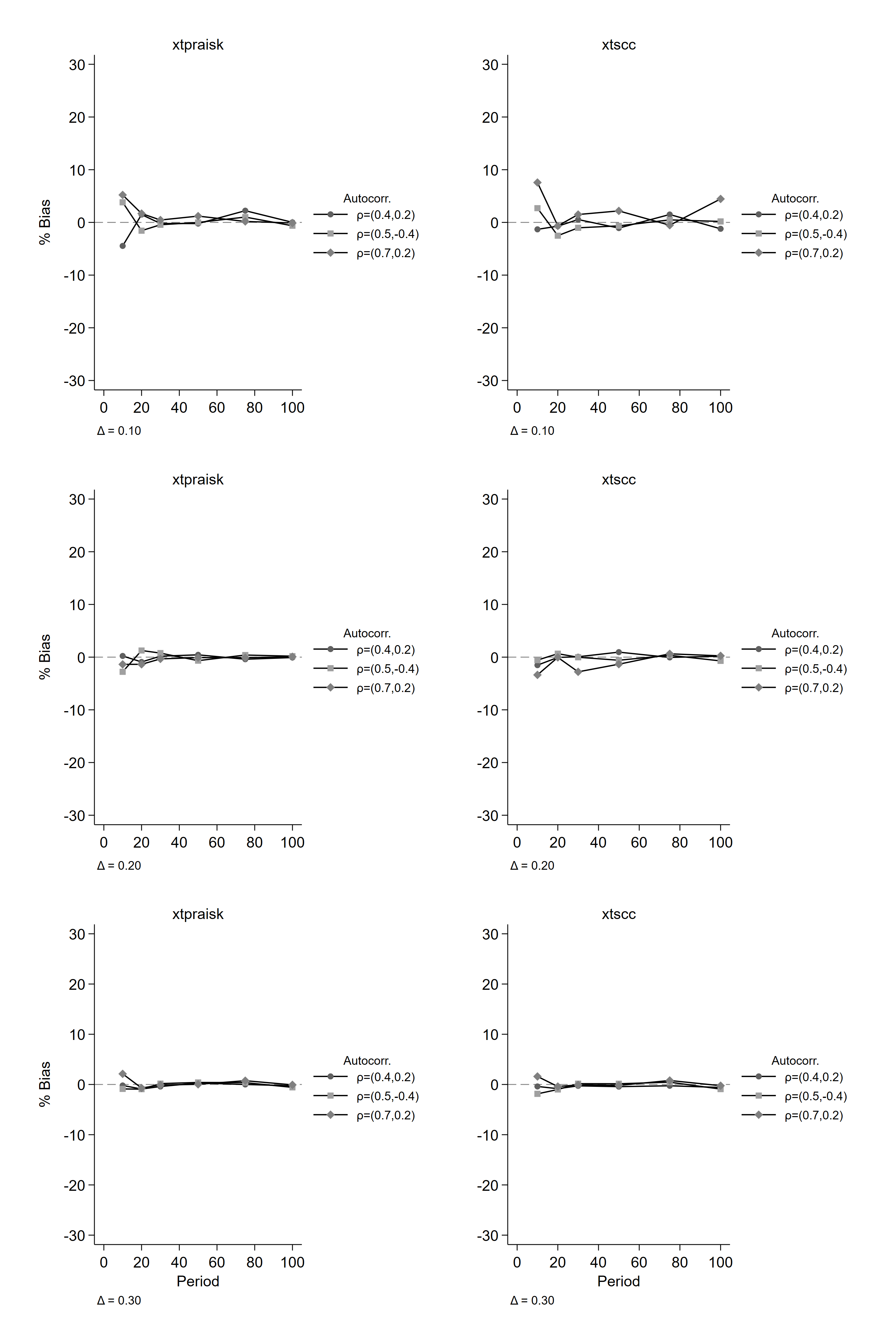}
\caption{Percentage bias for \texttt{xtpraisk} (left column) and \texttt{xtscc} (right column) under AR(2) error structures. Rows represent effect sizes ($\Delta = 0.10$, 0.20, 0.30). Lines distinguish autocorrelation scenarios: mild positive $\boldsymbol{\rho} = (0.4, 0.2)$ (circles); oscillatory $\boldsymbol{\rho} = (0.5, -0.4)$ (squares); high persistent $\boldsymbol{\rho} = (0.7, 0.2)$ (diamonds). Dashed reference line at zero. $N = 10$.}
\label{fig:bias_ar2}
\end{figure}

\clearpage
\begin{figure}[p]
\centering
\includegraphics[width=\textwidth,height=0.85\textheight,keepaspectratio]{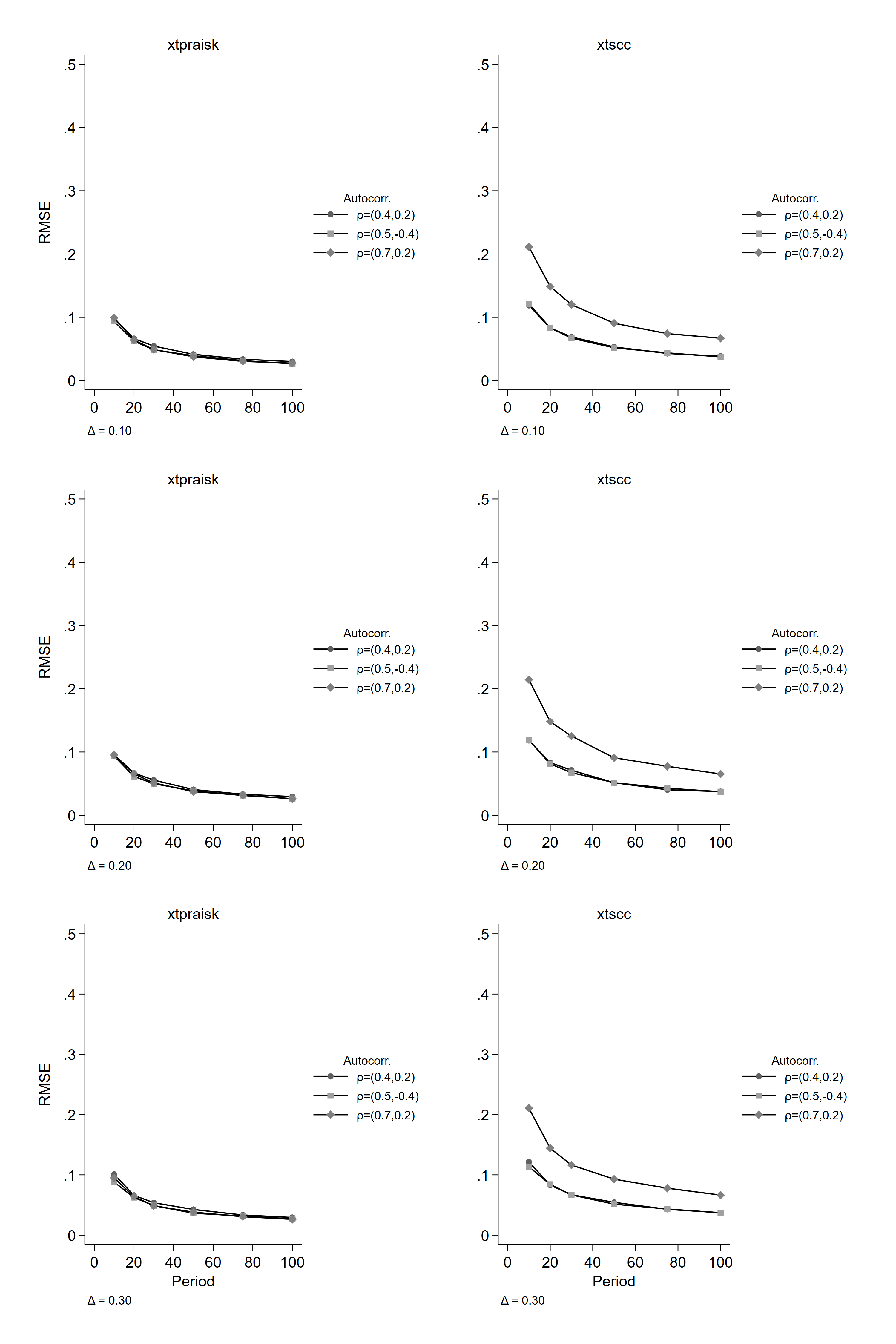}
\caption{Root mean squared error (RMSE) for \texttt{xtpraisk} (left column) and \texttt{xtscc} (right column) under AR(2) error structures. Rows represent effect sizes ($\Delta = 0.10$, 0.20, 0.30). Lines distinguish autocorrelation scenarios: mild positive $\boldsymbol{\rho} = (0.4, 0.2)$ (circles); oscillatory $\boldsymbol{\rho} = (0.5, -0.4)$ (squares); high persistent $\boldsymbol{\rho} = (0.7, 0.2)$ (diamonds). $N = 10$.}
\label{fig:rmse_ar2}
\end{figure}

\clearpage
\begin{figure}[p]
\centering
\includegraphics[width=\textwidth,height=0.85\textheight,keepaspectratio]{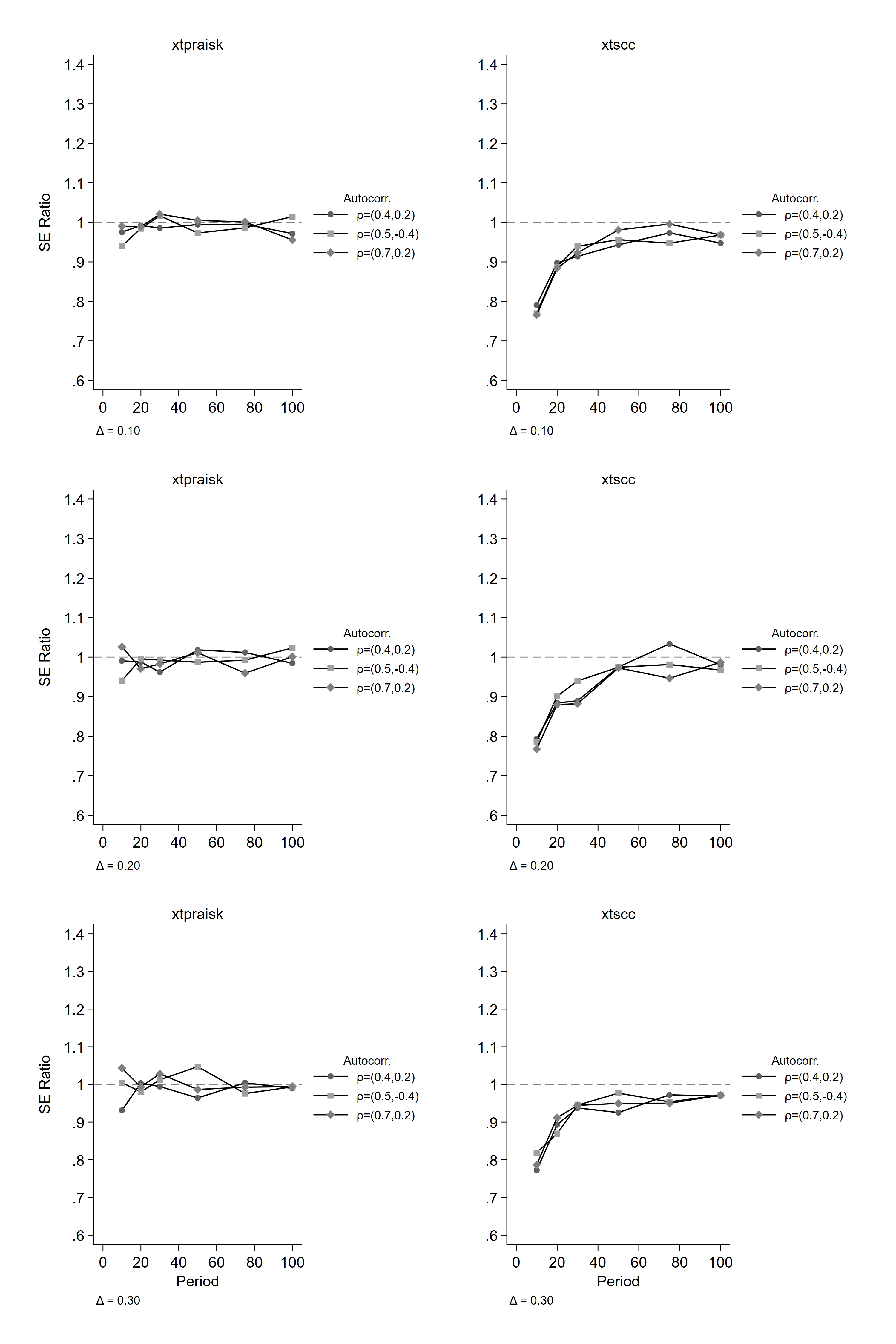}
\caption{Standard error (SE) ratio for \texttt{xtpraisk} (left column) and \texttt{xtscc} (right column) under AR(2) error structures. Rows represent effect sizes ($\Delta = 0.10$, 0.20, 0.30). Lines distinguish autocorrelation scenarios: mild positive $\boldsymbol{\rho} = (0.4, 0.2)$ (circles); oscillatory $\boldsymbol{\rho} = (0.5, -0.4)$ (squares); high persistent $\boldsymbol{\rho} = (0.7, 0.2)$ (diamonds). Dashed reference line at 1.0. Values below 1.0 indicate SE underestimation. $N = 10$.}
\label{fig:se_ar2}
\end{figure}

\clearpage
\begin{figure}[p]
\centering
\includegraphics[width=\textwidth,height=0.85\textheight,keepaspectratio]{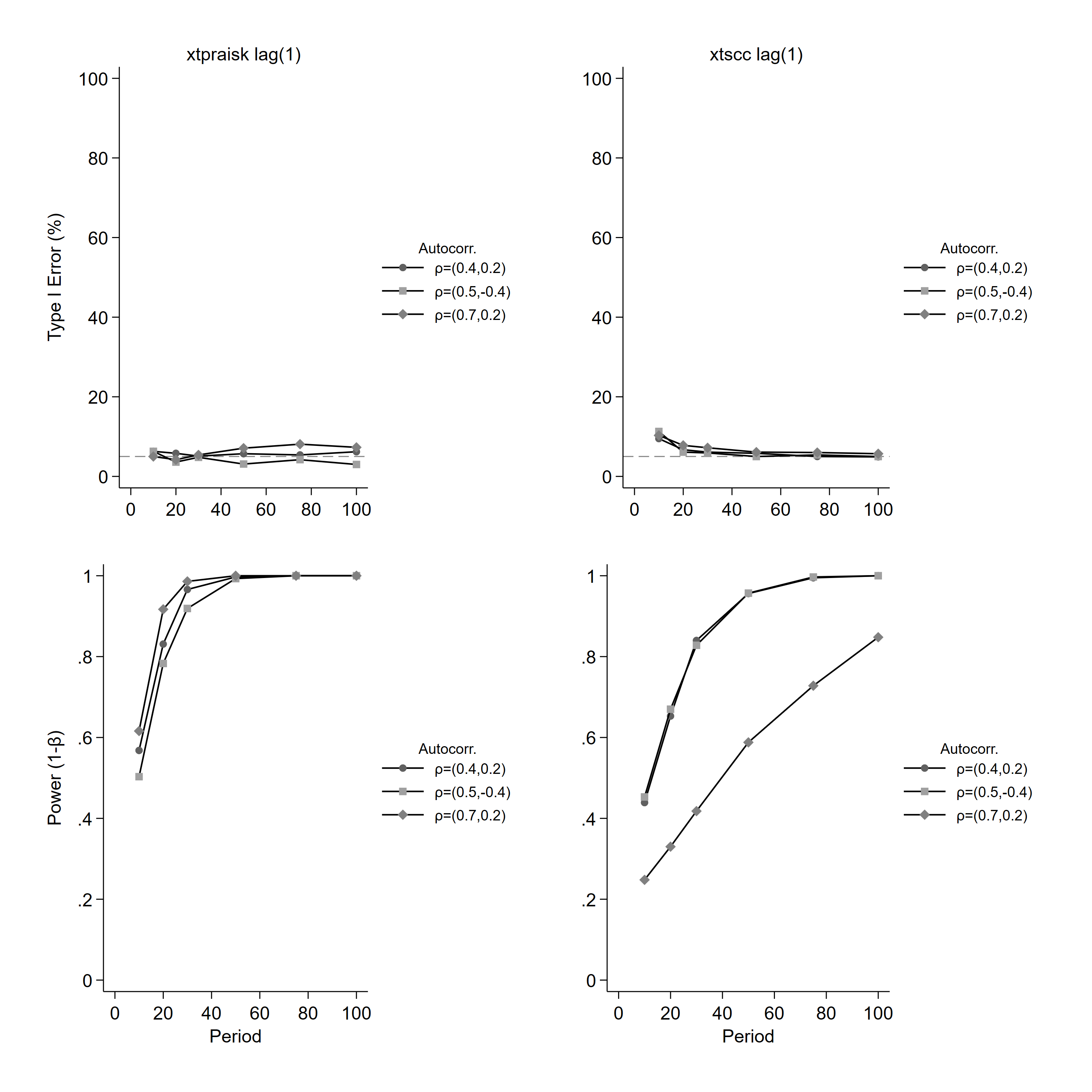}
\caption{Misspecification analysis: Type~I error (top row) and statistical power (bottom row) for \texttt{xtpraisk} lag(1) (left column) and \texttt{xtscc} lag(1) (right column) when data are generated under an AR(2) process but both estimators are fitted with a lag(1) model. Lines distinguish autocorrelation scenarios: mild positive $\boldsymbol{\rho} = (0.4, 0.2)$ (circles); oscillatory $\boldsymbol{\rho} = (0.5, -0.4)$ (squares); high persistent $\boldsymbol{\rho} = (0.7, 0.2)$ (diamonds). Dashed reference line at nominal 5\% (Type~I error). $N = 10$.}
\label{fig:misspec}
\end{figure}

\clearpage
\begin{figure}[p]
\centering
\includegraphics[width=\textwidth,height=0.85\textheight,keepaspectratio]{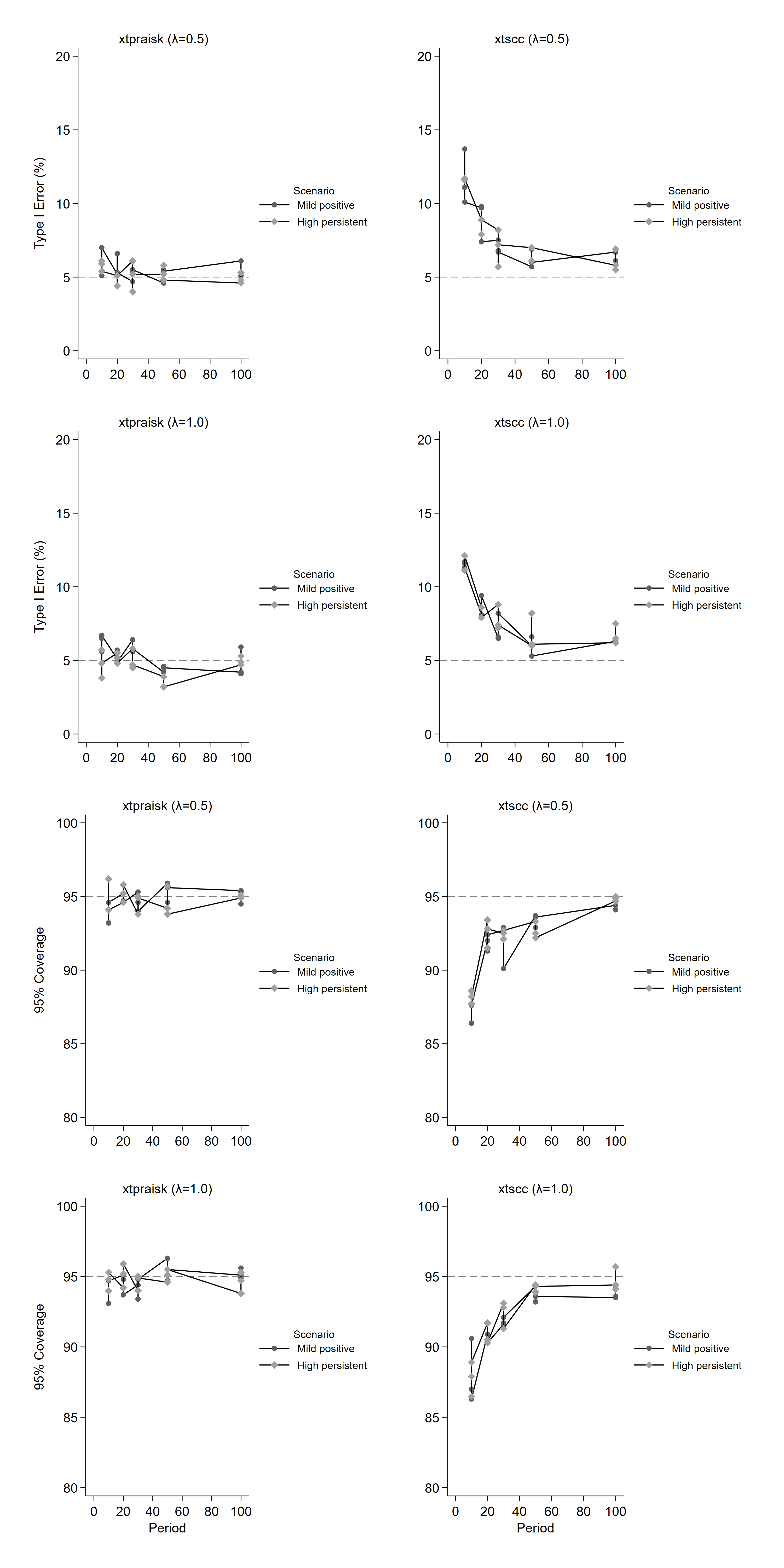}
\caption{Sensitivity analysis: Type~I error (top two rows) and 95\%
confidence interval coverage (bottom two rows) for \texttt{xtpraisk}
(left column) and \texttt{xtscc} (right column) under AR(2) errors with
a cross-panel common factor ($\lambda = 0.5$ and $\lambda = 1.0$). Lines
distinguish autocorrelation scenarios: mild positive
$\boldsymbol{\rho} = (0.4, 0.2)$ (circles); high persistent
$\boldsymbol{\rho} = (0.7, 0.2)$ (diamonds). Dashed reference lines at
nominal 5\% (Type~I error) and nominal 95\% (coverage). $N = 10$.}
\label{fig:sensitivity}
\end{figure}

\clearpage
\begin{figure}[p]
\centering
\includegraphics[width=\textwidth,height=0.85\textheight,keepaspectratio]{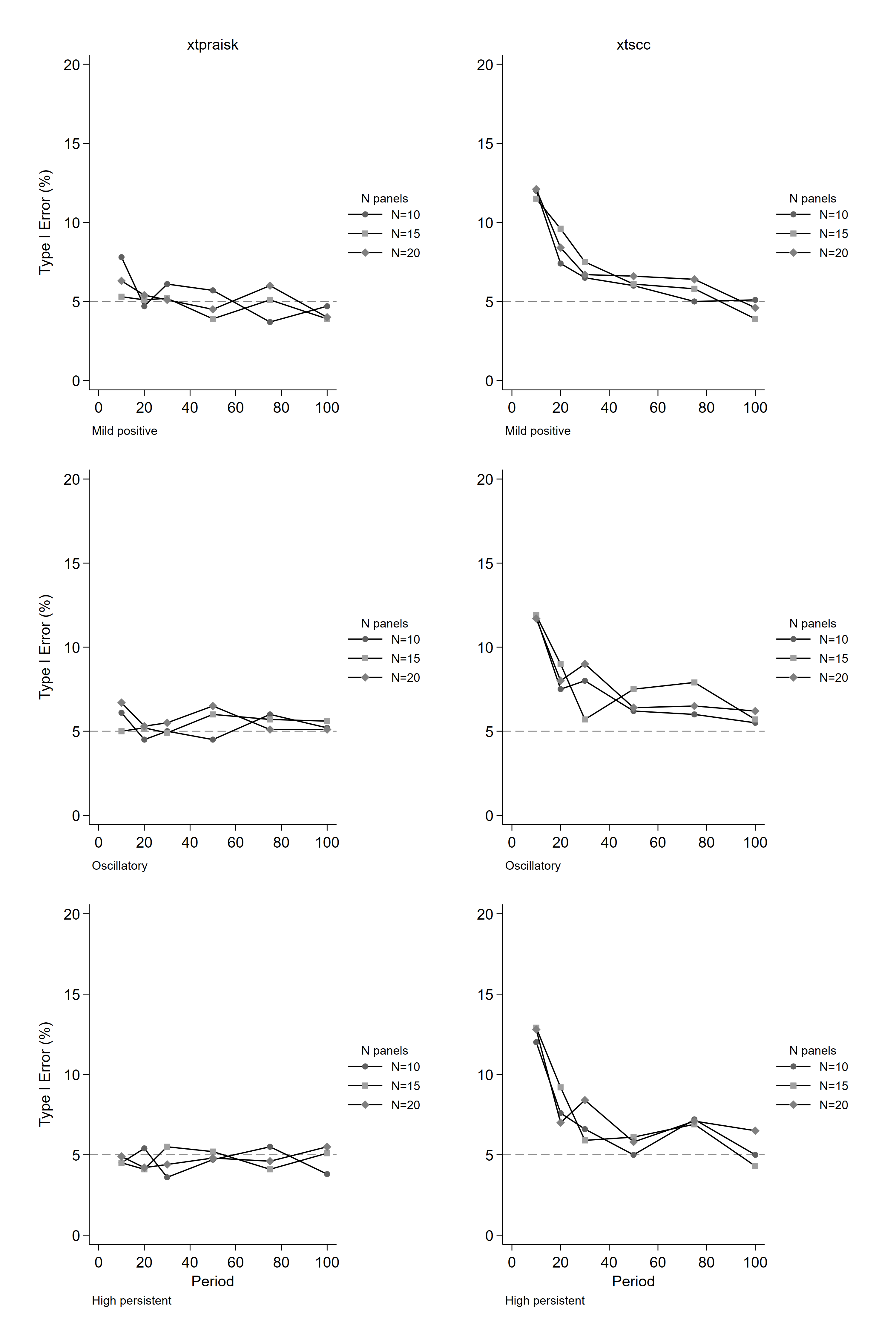}
\caption{Type~I error rates under AR(2) errors by panel size
($N = 10$, 15, 20). Left column: \texttt{xtpraisk}; right column:
\texttt{xtscc}. Rows represent autocorrelation scenarios: mild positive
$\boldsymbol{\rho} = (0.4, 0.2)$ (top); oscillatory
$\boldsymbol{\rho} = (0.5, -0.4)$ (middle); high persistent
$\boldsymbol{\rho} = (0.7, 0.2)$ (bottom). Lines distinguish panel
sizes: $N = 10$ (circles); $N = 15$ (squares); $N = 20$ (diamonds).
Dashed reference line at nominal 5\%. Effect size $\Delta = 0.20$.}
\label{fig:t1e_byN}
\end{figure}

\clearpage
\begin{figure}[p]
\centering
\includegraphics[width=\textwidth,height=0.85\textheight,keepaspectratio]{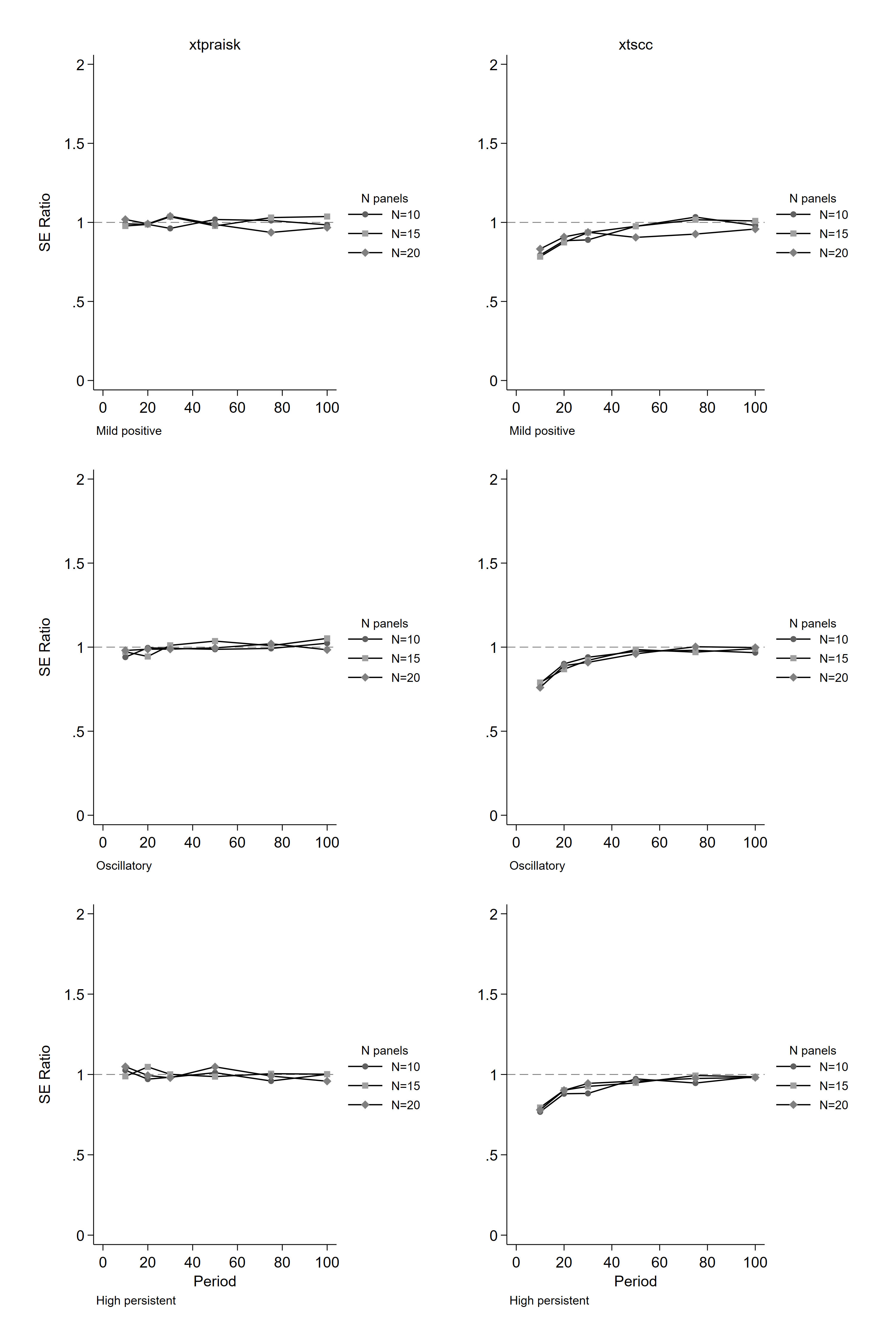}
\caption{Standard error (SE) ratio under AR(2) errors by panel size
($N = 10$, 15, 20). Left column: \texttt{xtpraisk}; right column:
\texttt{xtscc}. Rows represent autocorrelation scenarios: mild positive
$\boldsymbol{\rho} = (0.4, 0.2)$ (top); oscillatory
$\boldsymbol{\rho} = (0.5, -0.4)$ (middle); high persistent
$\boldsymbol{\rho} = (0.7, 0.2)$ (bottom). Lines distinguish panel
sizes: $N = 10$ (circles); $N = 15$ (squares); $N = 20$ (diamonds).
Dashed reference line at 1.0. Values below 1.0 indicate SE
underestimation. Effect size $\Delta = 0.20$.}
\label{fig:se_byN}
\end{figure}

\end{document}